\documentclass[showkeys,floatfix,amsfonts,amssymb,amsmath,aps,prb,reprint,superscriptaddress]{revtex4-2}
\usepackage[english]{babel}
\usepackage{tikz}
\usepackage{xcolor}
\usepackage{amsmath}
\newcommand{\order}[1]{\mathcal{O}(#1)}
\newcommand{\dd}[1]{\mathrm{d}#1}
\newcommand{\pdv}[2]{\dfrac{\partial #1}{\partial #2}}
\newcommand{\abs}[1]{|#1|}
\newcommand{\mel}[3]{\langle{#1}|{#2}|{#3}\rangle}
\newcommand{\ket}[1]{|#1\rangle}
\newcommand{\bra}[1]{\langle#1|}
\newcommand{\vdot}{\cdot}
\usepackage{siunitx}
\usepackage{graphicx}
\definecolor{color5}{HTML}{B0486D}
\definecolor{color4}{HTML}{B3A440}
\definecolor{color2}{HTML}{B0486D}
\definecolor{color1}{HTML}{379EB0}
\definecolor{color3}{HTML}{B3A440}
\newcommand{\hc}{\mathrm{H.~c.}}
\newcommand{\ncr}{\nonumber\\}
\newcommand{\y}[1]{_ {\phantom\dagger}^{#1}}
\newcommand{\w}[1]{^{\phantom\dagger}_{#1}}
\newcommand{\ed}{V\w{g}}

\newcommand{\ad}[1]{a^\dagger_{#1}}
\newcommand{\an}[1]{a_{#1}^{\phantom{1}}}

\newcommand{\cd}[1]{c^\dagger_{#1}}
\newcommand{\cn}[1]{c^{\phantom{\dagger}}_{#1}}
\newcommand{\dg}[1]{d^\dagger_{#1}}
\newcommand{\dn}[1]{d^{\phantom{\dagger}}_{#1}}

\newcommand{\fd}[1]{f^\dagger_{#1}}
\newcommand{\fn}[1]{f^{\phantom{\dagger}}_{#1}}
\newcommand{\bfd}[1]{{\bar f}^\dagger_{#1}}
\newcommand{\bfn}[1]{{\bar f}^{\phantom{\dagger}}_{#1}}

\newcommand{\mce}{\mathcal{E}}
\newcommand{\mcn}{\mathcal{N}}
\newcommand{\mcl}[1]{\mathcal{L}\w{#1}}
\newcommand{\mclu}[2]{\mathcal{L}^{#1}_{#2}}
\newcommand{\gd}[1]{g^\dagger_{#1}}
\newcommand{\gn}[1]{g^{\phantom{\dagger}}_{#1}}

\newcommand{\bn}[1]{b\w{#1}}

\newcommand{\ups}{\uparrow}
\newcommand{\dos}{\downarrow}
\newcommand{\kb}{k\w{B}}
\newcommand{\tk}{T\w{K}}
\newcommand{\vm}{\mathcal{V}}
\newcommand{\lm}{\mathcal{L}}

\newcommand{\hlm}{\mathcal{H}_{\mathrm{LM},\bar\delta}^{\ast}}
\newcommand{\hsc}{\mathcal{H}_{\mathrm{SC},\delta}^{\ast}}
\newcommand{\hkd}{\mathcal{H}\w{JW}}
\newcommand{\gb}[1]{\mathcal{G}\w{#1}}

\newcommand{\gtwo}{G\w{0}}
\newcommand{\uv}{E\w{\mbox{uv}}}
\newcommand{\offs}{\zeta}
\newcommand{\hdc}{H\w{dc}}
\newcommand{\himp}{H\w{dot}}
\newcommand{\bhimp}{\bar{H}\w{dot}}
\begin{document}
\title{Real-space numerical renormalization-group computation of transport properties in the side-coupled geometry}

\author{Ana Luiza Ferrari} \email[Correspondence email address:]{ana.luiza.ferrari@usp.br}
\affiliation{Institute for Theoretical Physics, University of Cologne, 50937 Cologne, Germany}
\affiliation{Instituto de Física de São Carlos, University of São 
  Paulo, 13560-970 São Carlos, SP, Brazil}

\author{Luiz N.~Oliveira}
\affiliation{Instituto de Física de São Carlos, University of São 
  Paulo, 13560-970 São Carlos, SP, Brazil}

\begin{abstract}
  The equilibrium transport properties of an elementary nanostructured
  device with side-coupled geometry are computed and related to
  universal functions. The computation relies on a real-space
  formulation of the numerical renormalization-group (NRG)
  procedure. The real-space construction, dubbed eNRG, is more straightforward than
  the NRG discretization and allows more faithful description of the
  coupling between quantum dots and conduction states. The procedure
  is applied to an Anderson-model description of a quantum wire
  side-coupled to a single quantum dot. A gate potential controls the
  dot occupation.  In the Kondo regime, the electrical conductance
  through this device is known to map linearly onto a universal
  function of the temperature scaled by the Kondo temperature. Here,
  the energy moments from which the Seebeck coefficient and the
  thermal conductance can be computed are shown to map linearly onto
  universal functions also. The moments and transport
  properties computed by the eNRG procedure are
  shown to agree very well with these analytical
  developments. Algorithms facilitating comparison with experimental
  results are discussed. As an illustration, one of the algorithms is
  applied to thermal dependence of the thermopower measured by
  K\"{o}hler [PhD Thesis, TUD, Dresden, 2007] in
  Lu$\w{0.9}$Yb$\w{0.1}$Rh$\w{2}$Si$\w{2}$.
\end{abstract}

\keywords{transport properties, NRG, side-coupled device, universality}

\maketitle
%%%%%%%%%%%%%%%%%

%section 1
\section{Introduction} \label{sec:introduction} The Numerical
Renormalization Group method was proposed five decades ago, to
calculate the thermodynamic properties of dilute magnetic alloys
\cite{Wilson,FirstUniversality,1980KWW1044}. Since then, the scope of
applications has been extended to include excitation and transport
properties \cite{ReviewNRG}. These developments and subsequent
advances have converted the method into an apt instrument in the
rapidly growing area of nano-device development \cite{2021PAP1863}.
Numerous examples constitute recent literature
\cite{2019DMCp506,2019Cos161106,2019Cos155126,2019XNDp102601,2020BBEp075132,2020BoF075149,2020YJZp085418,2020NDC115117,2020RiM241105,2020PSZp166259,2020TSHp165106,2020DDMp125115,2020EETp125405,2020MSRL095602,2021NLPp5878,2021DFS235166,2021PSV045137,2021ZOKp6320,2021ZaN035437}.

The modifications have made the procedure more efficient, more
accurate, or more general \cite{ReviewNRG}. All of them have
nonetheless preserved the core of Wilson's construction: logarithmic
discretization of the conduction band followed by a Lanczos
transformation, truncation, and definition of a renormalization-group
transformation.

That momentum-space construction seem less attractive today than it
was in the 1970's. Approximations in real space suit the geometry of
nanofabricated devices better than their counterparts in $k$-space. Of
course, Bloch states serve the most elementary designs well. Consider,
for example, the side-coupled device (SCD), a quantum dot side-coupled
to a quantum wire, or the single-electron transistor, a quantum dot
bridging two otherwise independent two-dimensional electron gases.
Momentum independent couplings models the tunneling between the
quantum dot and the conduction bands reliably. In
fact, rigorous renormalization-group arguments show that universal
properties are unaffected by the momentum dependence of the tunneling amplitudes
in simple arrangements \cite{1980KWW1044}.

In more elaborate geometries, by contrast, the momentum dependence may introduce marginal or even relevant operators \cite{1996SLOp275}. This raises an issue, for while the logarithmic discretization, the subsequent Lanczos transformation, and the truncation in the NRG construction leave momentum-independent couplings $\mathcal{V}$ intact, the same cannot be said of momentum-dependent couplings: the projection of the model Hamiltonian onto the basis of the logarithmically discretized states describes the momentum dependence only approximately. For typical discretization parameters, significant deviations are introduced.

Here, in an attempt to overcome this limitation of the NRG formalism, we present a real-space formulation of the method. The construction is analogous to the momentum-space formulation. Instead of lumping conduction states in logarithmically spaced intervals into discrete levels, the alternative approach assembles sites belonging to real-space blocks of exponentially growing size into discrete states. The resulting renormalization-group transformation is practically identical and preserves the virtues of the traditional approach: rapid convergence of physical properties to the continuum limit, uniform accuracy over parametrical spaces, access to the tools of renormalization-group theory, and relatively small computational cost.

As a reminder of exponential growth, we dub the alternative formulation \emph{eNRG}. In this report, instead of more complex geometries showing the full potential of the eNRG procedure, it is preferable to choose a simpler problem as an illustration, one in which numerical and analytical treatments can be dovetailed to corroborate preservation of the above-mentioned virtues. This considered, the object of our illustrative study will be the zero-bias transport properties of the side-coupled device.

The physical properties of the SCD are described by the Anderson Hamiltonian. In the Kondo regime, the thermal dependence of its conductance has been shown to map linearly onto a universal function of the temperature scaled by the Kondo temperature \cite{2009SYO67006,condutancia1}. The other SCD transport properties, the Seebeck or Peltier coefficients, and the thermal conductance, conform to no such mapping. They can, however, be computed from three energy moments. In analogy with a recent discussion of the SET geometry \cite{2021ARS085112}, we will show below that the pertinent energy moments map linearly to universal functions and that the mappings prove practical to interpret experimental data.

The presentation is split into nine sections. Section~\ref{sec:model} defines the system and model that will serve as test beds, and Sec.~\ref{sec:modif-numer-renorm} defines the eNRG construction and compares it with the NRG approach. To offer a preliminary view of results, Sec.~\ref{sec:comp-with-results} compares the exact temperature-dependent electrical conductance for a noninteracting model with the thermal dependences computed by the eNRG and NRG methods. Section~\ref{sec:energy-moments}, which describes the numerical procedure determining the transport properties, is followed by a discussion of universality, in Sec.~\ref{sec::universality}. Numerical results are presented in Sec.~\ref{sec:results}, while Sec.~\ref{sec:comp-with-exper} discusses comparison with experimental data. Section~\ref{sec:summary} summarizes the findings, followed by two appendices with technical details.

\section{Model}\label{sec:model}
\begin{figure}
 \includegraphics[width = 0.8\linewidth]{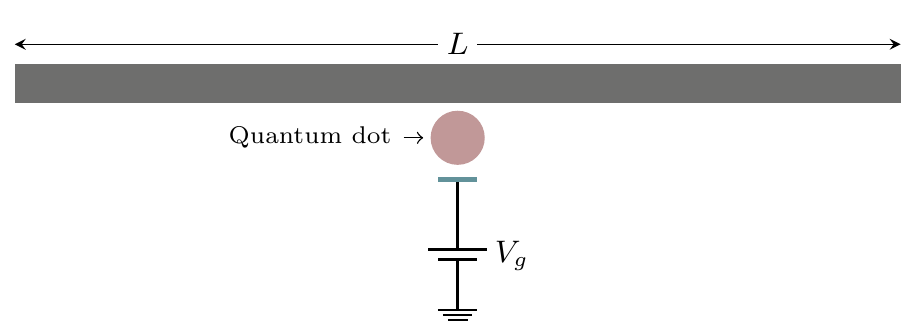}
 \caption[device]{\label{device} Side-coupled device. The quantum wire contains an odd number $L$ of sites. Electron tunneling is allowed between the quantum dot and the site at the center of the wire, which is assumed to have an odd number $L$ of sites. The gate potential $V\w{g}$ controls the dot occupation.}
 \end{figure}

As Fig.~\ref{device} indicates, the SCD comprises a quantum dot weakly coupled to a quantum wire.  Small electrical or thermal biases applied between the tips of the wire induce the electrical and thermal currents that determine the transport properties. At high temperatures, the coupling to the dot offers little resistance to conduction across the wire, even if the gate potential in the illustration is adjusted to favor formation of a magnetic moment at the quantum dot. Upon cooling, the coupling between the dot and conduction-electron spins gradually forms a Kondo cloud that screens the dot moment. The cloud obstructs transport. The electrical and thermal conductances are substantially reduced as the temperature falls below the Kondo temperature $\tk$.

 The single-impurity Anderson model captures the essential physics of the device. A state $\dn{}$ represents the quantum dot, and a conduction band, half-filled with noninteracting electrons, represents the quantum wire. The model Hamiltonian can be written in  the form
 \begin{align}\label{eq:1}
   H = H\w{cb} + \himp + \hdc.
 \end{align} 
 Here the first, second, and third terms on the right-hand side
 represent the wire, the quantum dot, and the coupling between them.

 Specifically, to describe the wire, we consider a lattice with $L$ sites and the tight-binding Hamiltonian:
\begin{equation}
    H\w{cb} = -t\sum\w{\ell \in \mathcal{L}} \left(\cd{\ell}\cn{l+1} + \hc \right),
    \label{eq:2}
\end{equation}
where spin sums are implicit, a notation followed throughout the paper, and $\lm = \{0, \pm1, \ldots, \pm(L-1)/2\}$.

The quantum dot is modeled by a single electronic orbital $d$, described by the Hamiltonian
\begin{equation}
    \himp =  V\w{g} \hat{n}\w{d}  + U n\w{d \uparrow} n\w{d \downarrow},
    \label{eq:3}
\end{equation}
where $V\w{g}$ is a gate potential, which controls the dot energy, and the term proportional to $U$ accounts for the Coulomb repulsion between the electrons in the doubly-occupied level. 

The last term on the right-hand side of Eq.~(\ref{eq:1}) is the coupling 
 \begin{equation}
\hdc =  \mathcal{V} \left(\dg{}\cn{0} + H.c.\right),
\label{eq:4}
\end{equation}
which models the tunneling between the dot orbital and the central wire site and hybridizes the dot with the conduction band. 

The hybridization broadens the dot level to the width
\begin{align}
  \label{eq:5}
  \Gamma = \pi\rho \mathcal{V}^2,
\end{align}
where
\begin{align}
  \label{eq:6}
  \rho = \dfrac{1}{2\pi t}
\end{align}
is the per-particle density of states at the Fermi level.

\subsection{Transformation to basis with well-defined parity}
\label{sec:transf-basis-with}
To simplify the numerical treatment of the model Hamiltonian, we exploit the left-right symmetry of the device in Fig.~\ref{device}. $H$ commutes with the parity operator $\Pi$, and its eigenstates can be classified by parity. The odd eigenstates are orthogonal to $\cn{0}$, the site that is directly coupled to the dot, and are hence decoupled from the dot.

Specifically, it is convenient to define the even operators
\begin{align}\label{eq:7}
  \an{0} &\equiv \cn{0},\\
  \an{\ell}& \equiv
             \dfrac{\cn{\ell}+\cn{-\ell}}{\sqrt2}\qquad(\ell=1,\ldots\bar L),\label{eq:8}
\end{align}
where $\bar L\equiv (L-1)/2$, and the odd operators
 \begin{align}\label{eq:9}
 \bn{\ell} \equiv \frac{\cn{\ell} -
   \cn{-\ell}}{\sqrt{2}}\qquad(\ell=1,\ldots, \bar L).
 \end{align}

On the basis of the $\an{\ell}$ and $\bn{\ell}$, the  Hamiltonian splits into an even and an odd term: $H = H\w{A} + H\w{B}$, where
\begin{align}\label{eq:10}
  H\w{A} =&  -\sqrt{2} t(a_{0}^{\dagger}a\w{1} + H.c.)
           -t\sum_{\ell=1}^{\bar{L}} (a^{\dagger}_{\ell}a\w{\ell+1}+\hc)\ncr
  &+ \himp + \hdc,
\end{align}
and
\begin{align}\label{eq:11}
    H\w{B} = -t\sum_{\ell=1}^{\bar{L}}\left(b^{\dagger}_{\ell}b\w{\ell+1} + H.c. \right).
\end{align}

Since the quadratic form~(\ref{eq:11}), which can be easily diagonalized, is completely decoupled from $H\w{A}$, we can focus our attention on the latter Hamiltonian, henceforth.

\subsection{Particle-hole transformation}\label{sec:part-hole-transf}
The conduction-band Hamiltonian and the coupling between the conduction band and the quantum dot, that is, the sum of the first, second, and last terms on the right-hand side of Eq.~(\ref{eq:10}), remain invariant under the particle-hole transformation
\begin{subequations}\label{eq:12}
\begin{align}
  d &\to -d^{\dagger},\label{eq:13}\\
  \an{\ell} &\to (-1)^{\ell} a^{\dagger}_{\ell}\qquad(\ell=1,\ldots, \bar L).\label{eq:14}
\end{align}
\end{subequations}

Application of~(\ref{eq:12}) to the dot Hamiltonian $\himp$ yields the following expression:
\begin{align}\label{eq:15}
    \bhimp =  -(U+V\w{g}) \hat{n}\w{d}  + U n\w{d\ups}n\w{d\dos} + U + 2V\w{g}.
\end{align}

The last two terms on the right-hand side constitute a constant, which merely redefines the energy zero. The particle-hole transformation maps $H\w{A}$ onto the conjugate Hamiltonian $\bar{H}\w{A}$, with the same model parameters, except for the gate potential $V\w{g}$, which undergoes the transformation
  \begin{align}\label{eq:16}
    V\w{g}\to -(U+V\w{g}).
  \end{align}

  For $V\w{g}= -U/2$ the two sides of Eq.~(\ref{eq:16}) become equal. This parametric choice defines the (particle-hole) symmetric model Hamiltonian.

\begin{figure}
  \includegraphics[width = 0.8\linewidth]{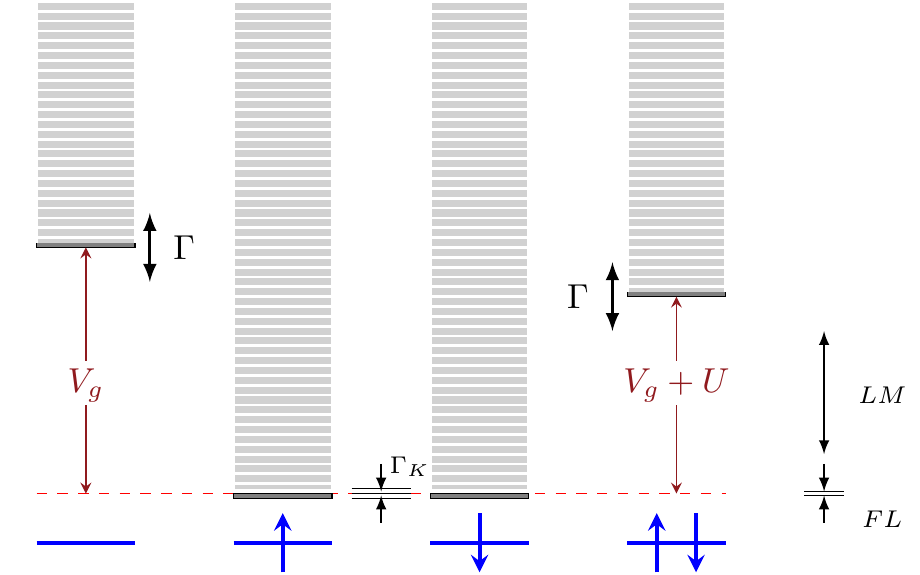}
    \label{QD_config}
    \caption[spectrum]{Spectrum of the Anderson model near the symmetric point. With no hybridization ($\Gamma=0$), the dot occupation $n\w{d}$ is conserved. The bold gray horizontal dashes represent the eigenvalues of the dot Hamiltonian, for the four configurations represented by the blue depictions at the bottom. The column with thin dashes above each dot energy represents conduction-band energies measured from the ground state.  The tunneling amplitude broadens the dot level to a width $\Gamma$ and hybridizes states in different columns. The arrows labeled $LM$ and $FL$ indicate fixed points of the renormalization-group transformation defined in the text. The arrows labeled $\Gamma\w{K}$ define the low-energy region of the spectrum, in which the dot and conduction electrons locks into a Kondo singlet.} 
\end{figure}

Fig. \ref{QD_config} represents the $\Gamma\to0$ spectrum of the model Hamiltonian in the vicinity of the symmetric point. The quantum dot is then decoupled from the conduction band, and the eigenvalues of the dot Hamiltonian~(\ref{eq:3}) are $\mathcal{E}\w{0}=0$, $\mathcal{E}\w{1\ups}=\mathcal{E}\w{1\dos}=\ed$, and
$\mathcal{E}\w{2}=2\ed+U$. Near particle-hole
symmetry, as the dark bold boxes in the figure indicate, the lowest
eigenvalue is degenerate,
$\mathcal{E}\w{min}=\mathcal{E}\w{1\ups}=\mathcal{E}\w{1\dos}$, and the energy
differences are approximately equal:
\begin{align}\label{eq:17}
\Delta\w{1}\equiv  \mathcal{E}\w{0}-\mathcal{E}\w{1} = -\ed, 
\end{align}
and
\begin{align}\label{eq:18}
\Delta\w{2}\equiv  \mathcal{E}\w{2}-\mathcal{E}\w{1} = \ed+U.
\end{align}

With $\Gamma=0$, each eigenstate of the model Hamiltonian is a combination of a many-body eigenstate of the conduction-band Hamiltonian with one of the four dot eigenstates, as indicated by the four stacks in the figure.

For $\max{\Delta\w{1},\Delta\w{2}}\gg \Gamma >0$, the coupling $\vm$ hybridizes states in the inner stacks with outside states. The coupling breaks the degeneracy between the spin-$1/2$ states in the central columns to form a singlet ground state separated from a triplet state by an energy of $\order{\Gamma\w{K}}$, where $\Gamma\w{K}\equiv \kb T\w{K}$ is the Kondo thermal energy. The hybridization between states in the two central stacks at energies $\epsilon\alt \kb T\w{K}$ defines the Kondo resonance. For temperatures below the Kondo temperature, represented by the energy interval $\Gamma\w{K}$ in the figure, the conduction band electrons screen the singly occupied impurity, forming the Kondo singlet.

The eigenvalues $\varepsilon$ of the Anderson Hamiltonian satisfying
$\varepsilon <\Gamma_{K}$ are beyond the reach of perturbation
theory. By contrast, the NRG describes accurately the entire spectrum
of the Hamiltonian, including this low-temperature region
\cite{Wilson}.
%section 1 
%%%%%%%

%section 2 
\section{Modified Numerical Renormalization Group method}
\label{sec:modif-numer-renorm}
The NRG method provides an efficient non-perturbative solution to quantum impurity systems. Even though it is historically associated with the Kondo problem, its applications are more general. It can be applied to systems where a quantum mechanical impurity is coupled to a non-interacting bath of fermions or bosons \cite{BosonicNRG}. There is
extensive literature on NRG concepts \cite{ReviewNRG}, and numerical implementations \cite{CodeNRG}. For this reason, this section presents an overview of the method, focused on aspects that distinguish the two approaches.

\subsection{NRG construction}
\label{sec:nrg-construction}
While the diagonalization of the odd term~\eqref{eq:11} of the model Hamiltonian is straightforward, the even term~\eqref{eq:10} requires numerical treatment. Brute-force diagonalization is possible for small lattices. The dimension of the Fock space grows exponentially with $L$, and this approach soon becomes unfeasible.

Alternatives are offered by the NRG and the eNRG approaches. The latter retains basic features of the former. Cursorily described, the two methods rely on strictly controllable approximations that project the conduction-band Hamiltonian upon discrete bases. The resulting discrete Hamiltonians are tight-binding forms with position-dependent
couplings and can be diagonalized iteratively with relatively small computational cost.

Notwithstanding the similarities, the two constructions are distinct. Brief recapitulation of the steps in Wilson's development seems therefore warranted, to facilitate comparison with the subsequent description of the real-space discretization.

\subsubsection{Logarithmic discretization of the conduction band}
\label{sec:logar-discr-cond}

The central approximation in the NRG procedure converts the conduction-band continuum to a logarithmic sequence of discrete levels. Specifically, Ref.~\cite{Wilson} considered a band with uniform density of states $\bar{\rho}$, half-filled with noninteracting electrons. The band Hamiltonian describes a continuum of spin-degenerate states $\cn{\epsilon}$ in the range $D\ge\epsilon\ge-D$: 
\begin{align}
  \label{eq:19}
  H\w{0} = \int_{-D}^D\epsilon \cd{\epsilon}\cn{\epsilon}\,\dd{\epsilon}.
\end{align}

The coupling between the impurity and the conduction band is given by the Hamiltonian
\begin{align}
  \label{eq:20}
  H\w{1} = \sqrt{2D}\mathcal{V}(\cd{d}\bfn{0}+\hc),
\end{align}
where 
\begin{align}
  \label{eq:21}
   \bfn{0} \equiv
  \sqrt{\dfrac{1}{2D}}\int_{-D}^{D}\cn{\epsilon}\,\dd{\epsilon}.
\end{align}

The operator $\bfn{0}$ defines a Wannier state with pivotal role in the NRG construction, because it controls the coupling between the conduction band and the impurity.

A dimensionless parameter $\Lambda>1$ defines the discretization of $H\w{0}$. Given $\Lambda$, the following expression introduces two infinite, discrete sequences of states:
\begin{align}
  \label{eq:22}
  \an{m\pm} \equiv \dfrac{1}{\mathcal{N}\w{m}}
  \int_{\pm D\Lambda^{-m-1}}^{\pm
  D\Lambda^{-m}}\cn{\epsilon}\,\dd{\epsilon}
  \qquad(m=0,1,\ldots),
\end{align}
where $\mathcal{N}\w{m}$ is a normalization factor.

The definition of the $\an{m\pm}$ offers an exact expression for the pivot:
\begin{align}
  \label{eq:23}
  \bfn{0} = \sqrt{\dfrac{1}{2D}}
  \sum\w{m,\eta=\pm} \mathcal{N}\w{m}\an{m\eta}.
\end{align}

The coupling between the dot level and the conduction states is not affected, therefore, by the discretization of the conduction band. This makes the NRG procedure uniformly accurate, for weak, moderate, or strong couplings.

While the coupling~\eqref{eq:20} can be faithfully described by the discrete operators, the conduction-band Hamiltonian cannot. The $\an{m\pm}$ constitute a basis that is incomplete relative to that of the $\cn{\epsilon}$. The substitution of the former basis for the latter one is justified \emph{a posteriori}, by the rapid convergence of computed physical properties to the continuum limit, as $\Lambda\to1$ \cite{Wilson,FirstUniversality}.

Projection of the conduction-band Hamiltonian upon the basis of discrete states, yields the approximate expression
\begin{align}
  \label{eq:24}
  H\w{0} = \sum_{m=0,\eta=\pm}^\infty\eta\mce\w{m}\ad{m\eta}\an{m\eta},
\end{align}
with the discrete energies
\begin{align}
  \label{eq:25}
  \mce\w{m} = D\dfrac{1+\Lambda^{-1}}2 \Lambda^{-m}.
\end{align}

\subsubsection{Conversion to a tridiagonal basis}
\label{sec:conv-trid-basis}

A Lanczos transformation starting with the operator $\bfn{0}$ in Eq.~(\ref{eq:21}) next converts the conduction-band Hamiltonian to the tridiagonal form
\begin{align}
  \label{eq:26}
H\w{0} = \sum_{n=0}^{\infty}\bar{t}\w{n} (\bfd{n}\bfn{n+1}+\hc),
\end{align}
where the $\bfn{n}$ are normalized Fermi operators, and the codiagonal coefficients have the expression 
\begin{align}
  \label{eq:27}
  \bar{t}\w{n} = D\dfrac{1+\Lambda^{-1}}{2}\dfrac{1-\Lambda^{-n-1}}
  {\sqrt{1-\Lambda^{-2n-1}}\sqrt{1-\Lambda^{-2n-3}}}\Lambda^{-n/2}.
\end{align}

For large $n$ the coefficients are accurately described by the simpler form
\begin{align}\label{eq:28}
  \bar{t}\w{n} = D\dfrac{1+\Lambda^{-1}}{2}\Lambda^{-n/2}\qquad(\Lambda^{-n}\ll 1).
\end{align}

The $\bfn{n}$ basis is complete with respect to the basis of the
$\an{m}$. The only approximation in the derivation of
Eq.~(\ref{eq:26}) is, therefore, the projection of the conduction-band
Hamiltonian upon the basis of the $\an{m}$. The parameter $\Lambda$ controls the accuracy of this approximation.

\subsubsection{Renormalization of coupling constants}
\label{sec:renorm-coupl-const}
To accelerate convergence to the continuum limit, it proved necessary
to renormalize the model parameters \cite{1980KWW1044}. Each operator
$\bfn{0}$ or $\bfd{0}$ in a model Hamiltonian must be multiplied by a
dimensionless factor $\sqrt{A\w{\Lambda}}$, where
\begin{align}
  \label{eq:29}
  A\w{\Lambda} = \dfrac{1+\Lambda^{-1}}{1-\Lambda^{-1}}\log\sqrt{\Lambda}
\end{align}
converges rapidly to unity as $\Lambda\to1$.

This upscaling is necessary because the discretization reduces the
$\bfn{0}$ spectral density by a factor $A\w{\Lambda}$. This
renormalization suffices to correct the procedure for
model Hamiltonians with energy-independent impurity-band couplings, such
as~(\ref{eq:20}). Energy-dependent couplings, such as those
in the two-impurity Anderson Hamiltonian, call for a
modified discretization procedure \cite{2005CaO104432}.

\subsubsection{Real-space approach}
\label{sec:real-space-numerical}
As explained in Sec.~\ref{sec:modif-numer-renorm}, the real-space
discretization procedure shares with traditional NRG 
the initial goal of reducing the conduction band Hamiltonian to a tridiagonal
form with progressively smaller off-diagonal coefficients. The
starting point, instead of Eq.~(\ref{eq:19}), is the tight-binding
Hamiltonian on the right-hand side of Eq. ~\eqref{eq:10}:
\begin{align}
  \label{eq:30}
  H\w{cb} &=  -\sqrt{2} t(a_{0}^{\dagger}a_{1} 
          + H.c.) -t\sum_{\ell =1}^{\bar L} (a^{\dagger}_{l}a_{\ell+1}
          + H.c.),
\end{align}
with $\bar L\to\infty$, which defines the continuum limit.

The discretization of the resulting spectrum is parametrized by two
natural numbers: the \emph{offset} $\offs\ge 1$, and
the \emph{common ratio} $\lambda$. The offset is a site index that divides the lattice into two sets: the first set comprises all sites to the left of site $\offs$; the second comprises the remaining sites. The discretization leaves intact those terms in the Hamiltonian associated with the first set, but affects those associated with the second set.

More specifically, the offset splits the right-hand side of Eq.~(\ref{eq:30}) into two tight-binding Hamiltonians:
\begin{align}\label{eq:31}
  H\w{cb} = H\w{a}+H\w{f},
\end{align}
where $H\w{a}$ comprises the first $\offs$ sites plus the coupling to site $\offs$:
\begin{align}\label{eq:32}
  H\w{a} \equiv-t\Big(\sqrt{2} a_{0}^{\dagger}a\w{1} +\sum_{\ell=1}^{\offs-1}
  a^{\dagger}_{\ell}a_{\ell+1} +\hc\Big), 
\end{align}
and $H\w{f}$ comprises the remaining sites
\begin{align}\label{eq:33}
  H\w{f} \equiv -t\sum_{\ell=\offs}^{\bar L}(\ad{\ell}\an{\ell+1} +\hc).
\end{align}

\begin{figure}
  \includegraphics[width = 0.8\linewidth]{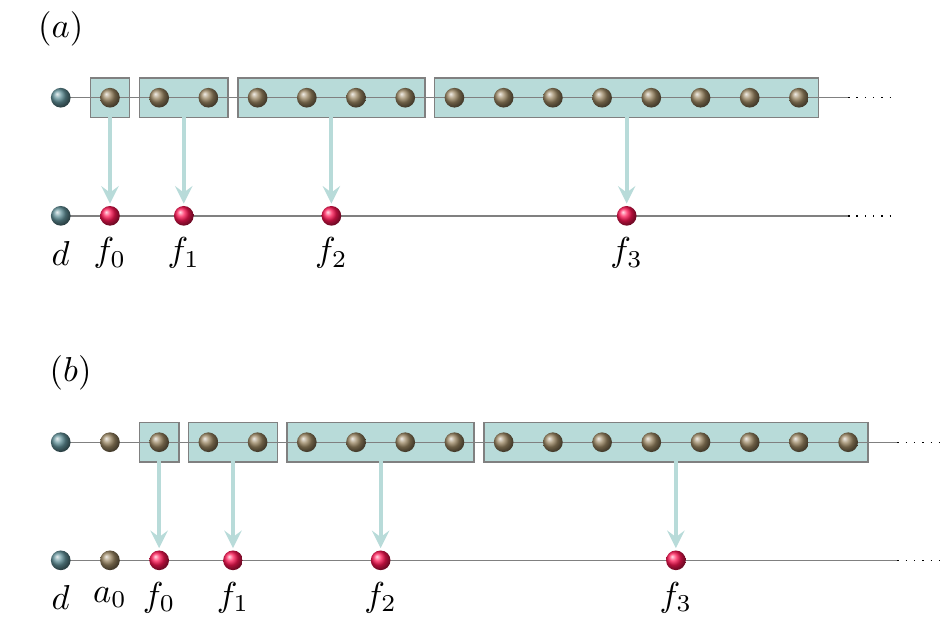}
     \caption[discrete]{\label{fig:discretization} Real-space geometry guiding the discretization of the conduction band, for \emph{common ratio} $\lambda=2$. The darker spheres at the top of each panel represent the dot and the wire lattice. The lattice sites are grouped into cells of exponentially increasing size. For each cell, Eq.~\eqref{eq:35} defines a linear combination $\fn{n}$ ($n=0, 1, \ldots$) of the wire states. The linear combinations form a basis $\{\fn{n}\}$ upon which the conduction-band Hamiltonian is projected.  The \emph{offset} $\offs$ specifies the position of the first cell. To illustrate, panels (a) and (b) depict the constructions for $\offs=0$ and $\offs=1$, respectively.}
 \end{figure}

The common ratio $\lambda$ defines a new basis comprising an infinite
set of Fermi operators $\fn{n}\qquad(n=0,1,2,\ldots)$, to
replace the operators $a\w{\ell}$ ($\ell=\offs,\ldots,\bar{L}$). As
Fig.~\ref{fig:discretization} indicates, the
definition starts out with an infinite sequence of cells $\mathcal{C}\w{n}$
($n=0,1,\ldots$). The first cell contains only one lattice site:
$j=\offs$. The second comprises $\lambda$ sites, from $\ell=\offs+1$ to
$\ell=\offs+\lambda$. 

The $n$-th cell covers $\lambda^n$ sites and extends from $\ell= \offs+ \gb{n}$ to $\ell= \offs + \gb{n} +\lambda^n$, where
\begin{align}
  \label{eq:34}
  \gb{n} \equiv \dfrac{\lambda^n-1}{\lambda-1}. 
\end{align}
is the geometric series with common ratio $\lambda$.

 With $\lambda=1$, each cell reduces to a single site, and we recover the continuum
 limit. With $\lambda >1$, the number of sites in cell
 $\mathcal{C}\w{n}$ grows exponentially with $n$. It is then
convenient to let $a\w{n,j}$ denote the Fermi operator
associated with the $j$-th site in cell $\mathcal{C}\w{n}$
($j=1,\ldots, \lambda^n$; $n=0,1,\ldots$), to avoid the cumbersome indexing
$a\w{\offs+\gb{n}+j-1}$.

A normalized linear combination $f\w{n}$ of the operators
$a\w{n,\ell}$ in cell $\mathcal{C}\w{n}$ can now be defined:
\begin{align}
  \label{eq:35}
  f\w{n} \equiv 
  \sum_ {j=1}^{\lambda^n}\alpha\w{n,j}a\w{n,j}\qquad(n=0,1,\ldots), 
\end{align}
with complex coefficients
\begin{align}
  \label{eq:36}
  \alpha\w{n,j} = |\alpha\w{n,j}|\exp(i\phi\w{n,j}),
\end{align}
which must satisfy the normalization condition:
\begin{align}
  \label{eq:37}
  \sum_{j=1}^{\lambda^n}|\alpha\w{n,j}|^2 = 1.
\end{align}

The definition (\ref{eq:35}) makes the operators $f\w{n}$ ($n=0,1,\ldots$) mutually orthogonal. The $f\w{n}$ form a basis that is incomplete with respect to the space spanned by the $a\w{n,j}$ ($n=0,1,\ldots$; $j=1,2,\ldots,\lambda^n$).  The incompleteness notwithstanding, following traditional NRG reasoning, we will project the conduction-band Hamiltonian onto the basis of the operators $d, a\w{0}, \ldots, a\w{\offs-1}, \fn{0}, \ldots, \fn{N-1}$. This approximation is justified \emph{a posteriori}, by the rapid convergence of physical properties of the $\lambda=1$ Hamiltonian.

Explicitly, the approximation amounts to treating Eq.~(\ref{eq:35})
as an orthonormal transformation, the inversion of which yields the
expression
\begin{align}
  \label{eq:38}
  a\w{n,j} =
  \alpha^*_{n,j}f\w{n}\qquad(n=0,1,\ldots; j=1,2,\ldots, \lambda^n). 
\end{align}

Substitution of the right-hand side of Eq.~(\ref{eq:38}) for the
$\an{\ell}$ expresses the Hamiltonian~(\ref{eq:33}) on the basis of
the operators $f\w{n}$ ($n=0,1,\ldots$):
\begin{align}
  \label{eq:39} 
  H\w{f\lambda} \equiv& -t\sum_{n=1}^{\infty}\sum_{j=1}^{\lambda^n}
                           (\alpha\w{n,j}\alpha_{n,j+1}^*+\mbox{c.~c.}) \fd{n}\fn{n}\ncr
&-t\sum_{n=0}^{\infty}(\alpha\w{n,\lambda\y{n}}\,\alpha_{n+1,1}^*\,\fd{n}\fn{n+1}+\hc). \end{align}

 The first term on the right-hand side in Eq.~(\ref{eq:39}) is at
 odds with Eq.~(\ref{eq:33}), which contains no diagonal
 terms. Judicious choice of the phases $\phi\w{n,j}$ is necessary and
 sufficient to eliminate this discrepancy.
 Appendix~\ref{sec:coeff-relat-oper} shows that the definition
 \begin{align}
   \label{eq:40}
   \phi\w{n,j} = \dfrac{\pi}2 (\gb{n}+n+j),
 \end{align}
reduces Eq.~(\ref{eq:39}) to the equality
 \begin{align}
   \label{eq:41}
  H\w{f\lambda} =
   t\sum_{n=0}^{\infty}(|\alpha\w{n,\lambda\y{n}}|\,|\alpha\w{n+1,1}|
   \,\fd{n}\fn{n+1}+\hc). 
 \end{align}

 If we now let the absolute values $|\alpha\w{n,j}|$ 
 ($n=0,1,\ldots$, $j=1,\ldots,\lambda^n$) be independent of $j$,
 the normalization condition~(\ref{eq:37}) yields
 \begin{align}
  \label{eq:42}
   |\alpha\w{n,j}| = \lambda^{-n/2},
 \end{align}
 which turns Eq.~(\ref{eq:41}) into an expression similar to Eq.~\eqref{eq:26}:

\begin{align}\label{eq:43}
  H\w{f\lambda} =   \sum_{n=0}^{\infty}(t\w{n}\fd{n}\fn{n+1}+\hc), 
 \end{align}
 where
 \begin{align}
  \label{eq:44}
   t\w{n} = t\lambda^{-n-\frac12}\qquad(n >0).
 \end{align}

 The proviso $n>0$ in Eq.~\eqref{eq:44} is necessary because $t\w{0}$ takes a special value if $\offs=0$. In that case, the operator $\fn{0}$ coincides with $\an{0}$, and the coupling between $\an{\offs}$ and $\an{\offs+1}$ on the right-hand side of Eq.~\eqref{eq:32} is $-\sqrt2t$, not $-t$. Hence
 \begin{align}
   \label{eq:45}
   t\w{0} =
   \begin{cases}
     t &\qquad(\offs=0)\\
     t\lambda^{-\frac12}&\qquad(\offs>0)
   \end{cases}.
 \end{align}
 
 Comparison between Eqs.~(\ref{eq:28})~and \eqref{eq:44} shows that, for $n$ such that $\Lambda^{-n}\ll 1$, the identification $\Lambda=\lambda^2$ brings the NRG codiagonal coefficients $\bar{t}\w{n}$ and the eNRG coefficients $t\w{n}$ into agreement, except for the constant prefactors $(1+\Lambda^{-1})/2$ and $\lambda^{-1/2}$. Although distinct, the two factors are approximately equal: they approach unity as $\Lambda, \lambda\to1$ and differ by less than 15\% for discretization parameters as large as $\Lambda=\lambda^2=25$.

 Substitution of $H\w{f\lambda}$ for $H\w{f}$ yields a discretized approximation to the Hamiltonian $H\w{A}$. Equation~\eqref{eq:10} becomes
 \begin{align}\label{eq:46}
  H\w{A\lambda} =& -t\Big(\sqrt{2}\ad{0} \an{1}
  +\sum_ {\ell=1}^{\offs-1}\ad{\ell}\an{\ell+1}+\hc\Big)\ncr
  &+ H\w{f\lambda}+\himp+\hdc.
\end{align}

The real-space construction can be regarded as a decimation procedure that spares sites closest to the quantum dot, but becomes rapaciously more inclusive as the  distance from the dot grows. The offset controls the size of the region in which all sites are spared and hence controls the eNRG resolution in the vicinity of the quantum dot. Larger $\offs$ offer more detailed description of the couplings.

Consider, for example, a quantum dot that is coupled to the central lattice site and to its nearest neighbors. The couplings are then described by the Hamiltonian
\begin{align}
  \label{eq:47}
  H_{dc}' = \Big(\vm \dg{}\an{0}+\mathcal{V}\w{1}\dg{}\an{1}+\hc\Big),
\end{align}
instead of $H\w{dc}$.

If $\offs=0$, the nearest neighbor will be one of two sites in cell $\mathcal{C}\w{1}$. The term $\mathcal{V}\w{1}(\dg{}\an{1}+\hc)$ will therefore be only approximately represented by the $\{\fn{n}\}$ basis, and the accuracy of the computation will depend on $\mathcal{V}\w{1}$.

The offset should instead be set to $\offs=1$ or larger. With $\offs=1$, the operator $\an{1}$ will coincide with $\fn{0}$ and the Hamiltonian $H_{dc}'$ will be exactly described on the basis of the $\fn{n}$.

Another aspect of this example deserves brief discussion. The addition of a coupling $(\mathcal{V}\w1\dg{}\an{1}+\hc)$ to the Hamiltonian~\eqref{eq:1} introduces only irrelevant operators, which affect such nonuniversal features of the single-impurity Anderson model as the Kondo temperature or the ground-state phase shift, but not its universal properties \cite{FirstUniversality}. This addition may break particle-hole symmetry, however, as one can check by applying transformation~\eqref{eq:12} to Eq.~\eqref{eq:47}.  Given that the physical properties of more complex Hamiltonians may depend critically on its symmetry \cite{1991Jon199153}, we can see that accurate description of the couplings to the wire may be necessary. Under these circumstances, the spatial resolution of the eNRG approach will be a valuable asset. 

\subsection{Truncation}
\label{sec:truncation}
Equation~(\ref{eq:43}) is closely analogous to the equality defining the logarithmically discretized conduction-band Hamiltonian in the standard NRG method.  This allows us to follow the truncation and iterative diagonalization procedure described in Ref.~\cite{FirstUniversality}.

The exponential decay on the right-hand side of Eq.~(\ref{eq:28}) allows definition of a renormalization-group transformation~\cite{Wilson}. To this end, consider an energy $\mathcal{E}$, representative of an energy scale of interest, and a dimensionless \emph{infrared-truncation} parameter $\gamma\ll 1$. One can then identify the smallest integer $\mcn$ satisfying the inequality \begin{align}
  \label{eq:48}
   t\lambda^{-\mcn} < \gamma \mathcal{E}.
\end{align}

Substitution of $\mcn-1$ for the upper limit of the sum then reduces the right-hand
side of Eq.~(\ref{eq:43}) to a finite series:
\begin{align}
  \label{eq:49}
  H\w{f\lambda} =
  \sum_{n=0}^{\mcn-1}t\w{n} (\fd{n}\fn{n+1}+\hc).
\end{align}

The inequality~(\ref{eq:48}) controls the accuracy of this
approximation. In the limit $\gamma\to0$, Eq.~(\ref{eq:49}) becomes
equivalent to Eq.~(\ref{eq:43}).

The right-hand side of Eq.~(\ref{eq:49}) can now be substituted for
$H\w{f\lambda}$ on the right-hand side of Eq.~(\ref{eq:46}). Next, the
resulting finite series is scaled up by the factor $1/t\w{\mcn-1}$,
which yields the dimensionless, truncated Hamiltonian $H\w{\mcn}$:
\begin{align}
  \label{eq:50}
  t\w{\mcn-1} H\w{\mcn} =&\himp
  +\hdc+\sum_{n=0}^{\mcn-1}t\w{n} (\fd{n}\fn{n+1}+\hc)\ncr
   & -t\Big(\sqrt{2}\ad{0} \an{1}+\sum_
             {\ell=1}^{\offs-1}\ad{\ell}\an{\ell+1}+\ad{\offs-1}\fn{0}
             +\hc\Big)
\end{align}

\subsection{Renormalization-group transformation and iterative
  diagonalization}
\label{sec:renorm-group-transf}

The truncation of the infinite series in the model Hamiltonian has
practical and conceptual implications. From the practical perspective,
the truncation is valuable because it allows iterative diagonalization of the
Hamiltonian, a procedure detailed in Ref.~\cite{CodeNRG}. At iteration
$\mcn$ ($n=0,1,\ldots$), the diagonalization determines all the
eigenvalues of $H\w{\mcn}$ below the \emph{ultraviolet cutoff}
$\uv$, a dimensionless parameter that controls the cost and the scope
of the diagonalization procedure. In addition, it gives access to the
matrix elements of the Fermi operators $\an{\ell}$
($\ell=0,\ldots,\offs-1$) and $\fn{n}$ ($n=0, \ldots,\mcn$) between
pairs of eigenvectors associated with the computed eigenvalues; the
effort necessary to determine such matrix elements is small in
comparison with the computational cost of diagonalizing the
Hamiltonian.

Conceptually, Eq.~(\ref{eq:50}) is important because it defines the
mapping $\tau[H\w{\mcn}] = H\w{\mcn+1}$, which adds a smaller
energy scale to $H\w{\mcn}$ and rescales the result so that the
resulting smallest eigenvalue be of $\order{1}$. The mapping is,
therefore, a renormalization-group transformation. From
Eq.~(\ref{eq:50}), it follows that
\begin{align}
  \label{eq:51}
  \tau[H\w{\mcn}] = \lambda H\w{\mcn} + (\fd{\mcn}\fn{\mcn+1}+\hc).
\end{align}

\subsection{Fixed points}
\label{sec:fixed-points}

As first discussed in Ref.~\cite{FirstUniversality}, for special 
combinations of the model parameters the Hamiltonian~$H\w{\mcn}$
is a fixed point of the renormalization-group transformation
$\tau^2$. Of special importance in this work are
(i) the \emph{local-moment} line of fixed points $\hlm$, and (ii)
the \emph{frozen-level} line of fixed points $\hsc$.

The vertical arrows labeled LM and FL at the extreme right in Fig.~\ref{QD_config}
indicate the energy ranges in which the spectrum of the model Hamiltonian is
close to the local-moment fixed point (LM) or the frozen-level fixed
point (FL). Near the LM, thermal or excitation energies are
higher than the energy scale $\kb T\w{K}$ defined by the Kondo
temperature $T\w{K}$; physically, the dot moment is free from
screening.

As the energy is reduced, the spectrum of the Hamiltonian
moves away from the LM structure, towards the FL structure. For energies
much smaller than the Kondo thermal energy, the dot moment is
completely screened, and the spectrum approaches the FL. 
 
Each local-moment fixed point is equivalent to a
phase-shifted conduction band decoupled from a spin-$1/2$ variable
$\vec S$ and is described by the truncated Hamiltonian
\begin{align}
  \label{eq:52}
  \hlm =
  \dfrac{1}{t\w{N-1}}\Bigg(\sum_{n=0}^{\mcn-1}t\w{n}(\fd{n}\fn{n+1}
  +\hc)+\bar W\fd{0}\fn{0}\Bigg), 
\end{align}
where the scattering potential $\bar W$ depends on the model parameters $U$, $\ed$, and $V$; for $\ed=-U/2$, in particular, particle-hole symmetry allows ony two potentials: $\bar W=0$ or $\bar W=\infty$.

The quadratic form on the right-hand side of Eq.~(\ref{eq:52}) can be diagonalized analytically \cite{1980KWW1044}. For definiteness, let $\mcn$ be odd. Then, there are $M\equiv (\mcn+1)/2$ positive eigenvalues $\bar\varepsilon\w{\ell+}$ and $M$ negative eigenvalues $\bar\varepsilon\w{\ell-}$, approximately given by the expressions
\begin{align}
  \label{eq:53}
  \bar\varepsilon\w{\ell\pm} = \pm\lambda^{2(\ell \mp\frac{\bar\delta}{\pi})}
  \qquad(\ell =0,1,\ldots,\dfrac{\mcn+1}2), 
\end{align}
with phase-shifts $\bar\delta$ determined by the scattering potential $\bar W$ on the right-hand side of Eq.~(\ref{eq:52}):
\begin{align}\label{eq:54}
  \bar \delta = -\arctan(\pi\rho\bar W). 
\end{align}
The phase shifts are defined $\mathrm{mod}\ \pi$, in the interval $-\pi/2<\bar\delta\le\pi/2$.

Projected onto the basis of its eigenvectors $\gn{\ell}$, the fixed-point Hamiltonian reads
\begin{align}
  \label{eq:55}
  \hlm = \sum\w{\ell,\alpha=\pm}\bar\varepsilon\w{\ell\alpha}
  \gd{\ell\alpha}\gn{\ell\alpha}.
\end{align}

The FL fixed points are equivalent to phase-shifted conduction bands, described by a Hamiltonian analogous to Eq.~(\ref{eq:52}):
\begin{align}
  \label{eq:56}
  \hsc =
  \dfrac{1}{t\w{N-1}}\Bigg(\sum_{n=0}^{\mcn-1}t\w{n}(\fd{n}\fn{n+1}
  +\hc)+W\fd{n}\fn{n}\Bigg).
\end{align}

The right-hand sides of Eq.~(\ref{eq:55})~and (\ref{eq:56}) have the
same form. The eigenvalues of $\hsc$ are therefore described by an
approximate equality analogous to Eq.~(\ref{eq:54}):
\begin{align}
  \label{eq:57}
  \varepsilon\w{\ell\pm} = \pm\lambda^{2(\ell \mp\frac{\delta}{\pi})}
  \qquad(\ell =0,1,\ldots,\dfrac{\mcn+1}2), 
\end{align}
where
\begin{align}
  \label{eq:58}
  \delta = -\arctan(\pi\rho W). 
\end{align}

Given that the spin of one electron is required to screen the dot moment,
the Friedel sum ties the two phase-shifts \cite{Friedel}:
\begin{align}
  \label{eq:59}
  \delta = \bar \delta -\dfrac{\pi}2.
\end{align}

A scattering potential $\bar W\ne0$ breaks particle-hole symmetry. For
the symmetric model, therefore, $\bar\delta=0$ and the frozen-level
phase shift is $\delta=\pi/2$.

\section{Comparison with the conductance for the uncorrelated Hamiltonian}
\label{sec:comp-with-results}

For $U=0$, the model Hamiltonian is quadratic. It is therefore
possible to diagonalize long chains, with very large $L'$. From the
resulting eigenvalues and eigenvectors, physical properties can be
accurately computed for energies $\mce$ much larger than the energy
splitting $\Delta E$ between successive single-particle levels in the
vicinity of the Fermi level. Comparison with the same properties
for the truncated Hamiltonian affords checks on the
accuracy of the approximations leading to Eq.~(\ref{eq:50}).

As an illustration, consider the thermal dependence of the electrical
conductance $G$. A simple expression for the conductance of the device
in Fig.~1 is available~\cite{condutancia1}; for $U=0$,
straightforward algebra reduces that expression to a sum involving
the single-particle eigenvalues and eigenvectors:
\begin{align}
  \label{eq:60}
  G(T) = \dfrac{\gtwo}{\rho}\sum_{n}\{\ad{0},\gn{n}\}^2
  \Bigg(-\pdv{f\w{\beta}(\epsilon)}{\epsilon}\Bigg)\w{\epsilon=\mce\w{n}},
\end{align}
where $\gtwo\equiv 2\mbox{e}^2/h$ is the conductance quantum, $\gn{n}$
and $\mce\w{n}$ denote the $n$-th single-particle eigenoperator and
the corresponding eigenvalue of the truncated
Hamiltonian~(\ref{eq:50}), respectively, $\beta\equiv 1/(\kb T)$, and
$f\w{\beta}( \epsilon)$ is the Fermi function
\begin{align}
  \label{eq:61}
  f\w{\beta}(\epsilon) = \dfrac1{1+\exp(\beta\epsilon)}.
\end{align}

\begin{figure}[!ht]
  \includegraphics[width=0.8\linewidth]{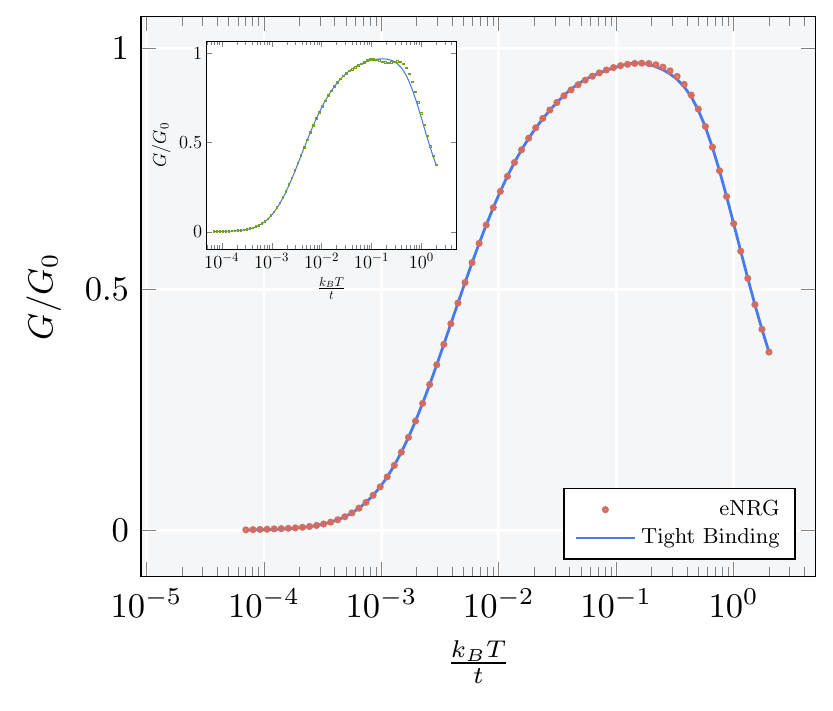}
  \caption[Noninteracting]{Thermal dependence of the conductance for
    the $U=\ed=0$ model, with level width $\Gamma=\num{1e-2}t$. The main plot shows the electrical conductance as a function of temperature. The circles were computed by the eNRG procedure, as described in the text; the solid line resulted from numerically diagonalizing the ($\lambda=1$) tight-binding Hamiltonian for $L=2001$. The inset displays the analogous comparison for the NRG procedure.}
  \label{fig:1}
\end{figure}

Figure~\ref{fig:1} shows conductances numerically computed from
Eq.~(\ref{eq:60}) for the $U=\ed=0$ model with
$\Gamma=\num{0.01}t$. The circles resulted from the
diagonalization of the Hamiltonian $H\w{\mcn}$ in Eq.~(\ref{eq:50}) with
$\lambda=2$ for two offsets: $\offs=3$ and $\offs=4$. Each point
is the arithmetic average between $G(\offs=3,T)$ and
$G(\offs=4,T)$.

The solid line in the figure represents conductances computed with
$\lambda=1$, that is, for the Hamiltonian~(\ref{eq:10}), with
$L'=10\,000$. The resultant level spacing near the Fermi level,
$\Delta E = 2\pi/L'$, allows accurate computation of the condutance
for temperatures above $\kb T=\num{1e-3}t$.

The inset displays conductances obtained from Eq.~(\ref{eq:60}) with
eigenvalues and eigenvectors resulting from NRG diagonalizations of
the model Hamiltonian, with the same model parameters. The circles
result from diagonalization with $\Lambda=4$ and $\mcn=40$, while
the solid line represents conductances computed with $\Lambda=1.0001$
and $\mcn=10\,000$. The solid lines in the main plot and inset are
pratically identical, even thoug the latter is based on the NRG
expression~(\ref{eq:19}), which describes a conduction band with linear
dispersion relation, while the former corresponds to a conduction band
described by the tight-binding Hamiltonian in Eq.~(\ref{eq:30}). 

The rapid decay of the conductance has simple physical
interpretation. At high temperatures, with $\kb T \gg \Gamma$, the
impurity is effectively decoupled from the conduction band and
allows ballistic transport through the chain. At low-temperatures,
with $\kb T\ll\Gamma$, the strong coupling to the impurity bars
transport across the site $\an{0}$ and reduces $G$ to zero.

As the plots show, at low temperatures both the NRG and eNRG
procedures yield essentially exact results. At higher temperatures,
the agreement between the circles and the solid line in the main plot
is also excellent, while the inset displays significant deviations for
$\kb T >0.1 t$. These discrepancies, of $\order{\kb T/D}$, are due to
irrelevant operators introduced by the logarithmic discretization of
the conduction band. By contrast, the eNRG procedure with offset
$\offs\ge 3$ describes the higher-energy degrees of freedom very well
and hence allows accurate computation of physical properties over the
entire temperature axis.

\section{Energy moments}
\label{sec:energy-moments}

Consider, now the transport properties for the correlated model. Of
special interest is the Kondo regime, defined by the inequality
\begin{align}
  \label{eq:62}
  |\ed+\dfrac{U}2| < \dfrac{U}2-\Gamma.
\end{align}

In the Kondo regime, as the temperature is lowered past $\min(|\ed|,
\ed +U)$, a local moment is formed at the quantum dot. Upon further
cooling, a Kondo cloud gradually screens the magnetic moment. Well
below the characteristic Kondo temperature $T\w{K}$, the physical
properties associated with the quantum dot show that the
effective magnetic moment has been reduced to zero.

In the same way
that the strong coupling between the quantum dot and the $\an{0}$
orbital blocks electrical conductance at low temperatures in
Fig.~\ref{fig:1}, the formation of the Kondo cloud affects the
transport properties of the side-coupled device. This section discusses
the computation of three temperature-dependent energy moments, from
which the electrical and thermal conductances, and the thermopower can
be obtained.

Specifically, the following three moments have to be computed:
\begin{equation}
    \mcl{j} \equiv \frac{2}{\rho h} \int \left(-\frac{\partial
        f\w{\beta}}{\partial \epsilon} \right) (\beta \epsilon)^{j}
    \rho\w{0}(\epsilon, T) d\epsilon\qquad(j=0,1,2), 
\label{eq:63}
\end{equation}
where $\rho\w{0}(\epsilon, T)$ is the spectral density of the $\an{0}$ orbital:
\begin{equation}
    \begin{aligned}
      \rho\w{0}(\epsilon, T) =& \frac{1}{\mathcal{Z}f\w{\beta}(\epsilon)}
      \sum\w{m,n} e^{-\beta E_{m}}
      \abs{\mel{m}{\ad{0}}{n}}^{2} \delta(\epsilon - E\w{mn}).
    \end{aligned}
    \label{eq:64}
\end{equation}
Here $\ket{p}$ and $E\w{p}$ ($p=m,n$) denote the $p$-th eigenvector and the corresponding eigenvalue of the model Hamiltonian, respectively, $E\w{mn}\equiv E\w{m}-E\w{n}$, and $\mathcal{Z}(T)$ is the partition function.

Substitution of the right-hand side of Eq.~(\ref{eq:64}) for the spectral density on the right-hand side of Eq.~(\ref{eq:63}) yields a simpler expression:
\begin{align}
  \label{eq:65}
    \mcl{j} = \dfrac{2}{h}
  \frac{\beta}{\rho\mathcal{Z}}\sum\w{m,n}\dfrac{\abs{\mel{m}{\an{0}}{n}}^{2}}{e^{\beta
  E\w{m}}+e^{\beta E\w{n}}}(\beta E\w{mn})^{j} \qquad(j=0,1,2).
\end{align}

Once the energy moments are computed, the following equalities yield the electrical conductance $G$, thermopower $S$, and thermal conductance $\kappa$ \cite{yoshida2009thermoelectric,kim2002thermopower}: 
\begin{equation}\label{eq:66}
    G(T) = e^{2} \mcl{0}(T);
\end{equation}
\begin{equation}\label{eq:67}
    S(T) = - \frac{\mcl{1}(T)}{e\mcl{0}(T)};
  \end{equation}
  and
\begin{equation}
    \beta \kappa(T) = \mcl{2}(T) - \dfrac{\mathcal{L}^{2}_{1}(T)}{\mcl{0}(T)}. 
    \label{eq:68}
\end{equation}

In practice, to determine the transport properties we only have to compute the three moments $\mcl{j}$ ($j=0,1,2$). This is a relatively simple task, since the iterative diagonalization of the truncated Hamiltonian~(\ref{eq:50}) determines the eigenvalues, and the recursive procedure introduced in Ref.~\cite{CodeNRG} gives immediate access to the matrix elements on the right-hand side of Eq.~(\ref{eq:65}).

\section{Universality}\label{sec::universality}

Along with the introduction of the NRG method, Wilson offered a renormalization-group analysis proving the crossover from the LM to the FLs to be universal \cite{Wilson}; the thermodynamical properties for the Kondo model and Kondo regime of the Anderson model are universal functions of the temperature scaled by the Kondo temperature \cite{Wilson,FirstUniversality}. Zero-temperature excitation properties are, likewise, universal functions of the energy scaled by the Kondo thermal energy \cite{1986FrO7871}.

Besides providing insight and simplifying theoretical analyses,
universality aids the interpetation of experimental results. Fits to
laboratory data offer \emph{prima-facie} evidence of Kondo physics
\cite{1998GSM+156}, for instance.

In contrast with the temperature dependence of thermodynamical
properties and with the frequency dependence of excitation properties,
the temperature-dependent transport properties cannot be universal
functions of $T/T\w{K}$. Straightforward scattering theory analysis
shows that, at the LM, the probability for transmission
across the side-coupled device is $\cos^2\bar\delta$. The three energy
moments must, therefore, depend on $\bar\delta$, which is
model-parameter dependent.

For the symmetric model, however, renormalization-group analysis showed that the Kondo-regime electrical conductance of the side-coupled device is a universal function $G\w{\mbox{uni}}(T/T\w{K})$, of the temperature scaled by the Kondo temperature. For asymmetric models, the conductance maps linearly onto $G\w{\mbox{uni}}$, with coefficients dependent on the phase shift $\delta$ only \cite{2009SYO67006,condutancia1}. Recent work has extended the approach to the three energy moments that determine the transport properties for the single-electron transistor \cite{2021ARS085112}. It resulted that, for the symmetric model, each moment is a universal function $f\w{n}(T/T\w{K})$ and that, for asymmetric models, the $n$-th moment maps linearly onto $f\w{n}(T/T\w{K})$, with $n$-dependent coefficients fixed by the FL. This section extends the findings of Refs.~\cite{2009SYO67006,condutancia1} to the energy moments in Eq.~(\ref{eq:63}). 

\subsection{Universal matrix elements} \label{sec:univ-matr-elem}
In the Kondo regime, well above the Kondo temperature, the model Hamiltonian lies close to the LM. The deviations are described by the Kondo Hamiltonian $\hkd$, with parameters determined by the Schrieffer-Wolff transformation \cite{citeSW}: \begin{equation}
    \hkd = \sum\w{k} \epsilon_{k} a_{k}^{\dagger}a_{k} + \bar W a^{\dagger}_{0}a_{0} + J \sum_{\mu\nu} a_{0\mu}^{\dagger} \mathbf{\sigma}_{\mu\nu} a_{0\nu} \cdot \mathbf{S},
    \label{eq:69}
\end{equation}
where
\begin{equation}\label{eq:70}
   \rho  J =  \dfrac{2\Gamma U}{\pi\abs{\ed}(\ed+U)},
  \end{equation}
  and
  \begin{align}\label{eq:71}
    \rho \bar W = \dfrac{\Gamma(\ed+\dfrac{U}2)}{\pi\abs{\ed}(\ed+U)}.
  \end{align}

  Combined with Eq.~(\ref{eq:71}), Eq.~(\ref{eq:54}) yields the LM
  phase shift $\bar\delta$ and, through Eq.~(\ref{eq:59}), the FL
  phase shift $\delta$. The following analysis shows that this is
  sufficient to determine the thermal dependence of the transport
  properties.

  To start out, we must eliminate the scattering potential from Eq.~\eqref{eq:69}. To this end, it is sufficient to project $\hkd$ onto the basis of the $\hlm$ eigenvectors $\gn{\ell}$. The
  projection generates irrelevant operators, which can be safely
  dropped. The remaining terms yield the expression \cite{YSO2009:235317}
\begin{equation}
    \hkd = \sum_{\ell} \varepsilon_{\ell} g_{\ell}^{\dagger} g_{\ell}
    + J\w{\bar W} \sum\w{\mu \nu} \phi_{0\mu}^{\dagger}
   \vec{\sigma}\w{\mu\nu} \phi\w{0\nu} \vdot \vec{S}
    \label{eq:72}
\end{equation}
where
\begin{equation}
  J\w{\bar W} = J \cos^{2}(\bar\delta),
\end{equation}
and
\begin{align}
  \label{eq:73}
  \phi\w{0} \equiv \dfrac1{\sqrt{N}}\sum\w{\ell}\gn{\ell}. 
\end{align}

At high energies, comparable to the conduction bandwidth $4t$, the
 contribution from the irrelevant operators makes the spectra of the
 Hamiltonians on the right-hand sides of
 Eqs.~(\ref{eq:69})~and~(\ref{eq:72})
 somewhat different. As the energy $\mce$ decreases, however, the deviations
 shrink and become negligible for $\mce\ll t$.

 Equation~(\ref{eq:72}) describes the physical properties of the model Hamiltonian in the vicinity of the LM. As the energy scale is reduced, the eigenstates and eigenvalues of the model Hamiltonian progressively deviate from the spectrum of $\hlm$. It is then convenient to switch to another basis, in which the basis vectors are linear combinations of the operators $\gn{\ell}$ with the Legendre
 polynomials $P\w{k}(\epsilon)$ ($k=0,1,\ldots$) as coefficients:
 \begin{align}
   \label{eq:74}
   \phi\w{k} \equiv \mcn\w{k}\sum\w{\ell} P\w{k}(\epsilon\w{\ell})\gn{\ell}\qquad(k=0,1,\ldots),
 \end{align}
 with appropriate normalization factors $\mathcal{N}\w{k}$.

 The leading basis vector is $\phi\w{0}$, defined by Eq.~(\ref{eq:73}).
Next comes the operator
\begin{align}\label{eq:75}
  \phi\w{1} \equiv \sqrt{\dfrac{\lambda^2-1}{2\lambda}}
  \sum\w{\ell}P\w{1}(\epsilon\w{\ell})g\w{\ell},
\end{align}
which plays an important role in the following analysis.

The second term on the right-hand side of
Eq.~(\ref{eq:72}) is a marginally relevant operator,
which brings the Hamiltonian from the LM to the FL. The
trajectory in renormalization-group space is universal, since a single
operator drives the flow. The coupling $J\w{\bar W}$ defines the
Kondo temperature $\tk$.  Scaling by $\kb\tk$ brings the
spectra of Hamiltonians of the form~(\ref{eq:72}) with different
couplings to congruence.

In other words, the same eigenvalues and eigenstates contribute to
physical properties computed for Hamiltonians with different couplings
at temperatures such that the ratio $T/T\w{K}$ is the same. 
Matrix elements of the operators $\phi\w{0}$ and $\phi\w{1}$
between eigenstates of the Kondo Hamiltonian are likewise universal.

To determine the energy moments~(\ref{eq:65}), the matrix elements
$\mel{m}{\an{0}}{n}$ must be computed, where $\ket{m}$ and $\ket{n}$
are eigenstates of the truncated Hamiltonian~(\ref{eq:50}).  The
matrix elements are linear combinations of the matrix elements
$\mel{m}{\phi\w{0}}{n}$ and $\mel{m}{\phi\w{1}}{n}$
\cite{YSO2009:235317}:
\begin{align}
  \label{eq:76}
  \mel{m}{\an{0}}{n} = \alpha\w{0}\mel{m}{\phi\w{0}}{n}+\alpha\w{1}\mel{m}{\phi\w{1}}{n}, 
\end{align}
with model-parameter dependent coefficients $\alpha\w{0}$ and
$\alpha\w{1}$.

Given that all coefficients $t\w{n}$ in the truncated Hamiltonian are real, the $\alpha\w{n}$ ($n=0,1$) and matrix elements in Eq.~(\ref{eq:76}) can be asssumed real, with no loss of generality. The squared matrix element in the summand on the right-hand side of Eq.~(\ref{eq:65}) are therefore given by the equality
\begin{align}
  \label{eq:77}
  \abs{\mel{m}{\an{0}}{n}}^2 =&
  \alpha_{0}^2\mel{m}{\phi\w{0}}{n}^2
    +\alpha_{1}^2\mel{m}{\phi\w{1}}{n}^2\ncr
  &+2\alpha\w{0}\alpha\w{1}\mel{m}{\phi\w{0}}{n}\mel{m}{\phi\w{1}}{n}.
\end{align}

Substitution of the right-hand side for $\abs{\mel{m}{\an{0}}{n}}^2$ in
  Eq.~(\ref{eq:65}) splits each energy moment into three terms:
\begin{align}
  \label{eq:78}
  \mcl{j} = \alpha_0^2\mclu{j}{00} +
  2\alpha\w{0}\alpha\w{1}\mclu{j}{01} +\alpha_1^2\mclu{j}{11}\qquad(j=0,1,2),
\end{align}
where
\begin{align}
\label{eq:79}
    \mclu{j}{00} \equiv \dfrac{2}{\rho h}
  \frac{1}{\mathcal{Z}}\sum\w{m,n}\dfrac{\bra{m}\phi\w{0}\ket{n}^{2}}{e^{\beta
  E\w{m}}+e^{\beta E\w{n}}}(\beta E\w{mn})^{j};
\end{align}
\begin{align}
\label{eq:80}
    \mclu{j}{01} \equiv \dfrac{2}{\rho h}
  \dfrac{1}{\mathcal{Z}}\sum\w{m,n}&\dfrac{\mel{m}{\phi\w{0}}{n}\mel{m}{\phi\w{1}}{n}}
  {e^{\beta E\w{m}}+e^{\beta E\w{n}}}(\beta E\w{mn})^{j};
\end{align}
and
\begin{align}
  \label{eq:81}
  \mclu{j}{11} \equiv \dfrac{2}{\rho h}
  \frac{1}{\mathcal{Z}}\sum\w{m,n}\dfrac{\bra{m}\phi\w{1}\ket{n}^{2}}
  {e^{\beta  E\w{m}}+e^{\beta E\w{n}}}(\beta E\w{mn})^{j}.
\end{align}

The summand on the right-hand side of Eq.~(\ref{eq:74}), which defines
$\phi\w{k}$ ($k=0,1,\ldots$), is proportional to the Legendre
Polynomial $P\w{k}(\epsilon)$, a function of the energy with the
parity of $k$. The matrix elements $\mel{m}{\phi\w{k}}{n}$ and
$\mel{n}{\phi\w{k}}{m}$ hence have the same sign for $k=0$ and
opposite signs for $k=1$.  For even $j$, therefore, the summands on
the right-hand sides of Eqs.~(\ref{eq:79})~and (\ref{eq:81}) remain
invariant under exchange of the summation indices ($m\leftrightarrow n$), while
the summand in Eq.~(\ref{eq:80}) changes sign. Conversely, for odd $j$
the summands in Eqs.~(\ref{eq:79})~and (\ref{eq:81}) change sign under
index exchange, while the summand on the right-hand side of
Eq.~(\ref{eq:80}) remains invariant. It follows that
\begin{align}
  \label{eq:82}
  \mclu{0}{01} = \mclu{2}{01} = \mclu{1}{00} = \mclu{1}{11} = 0.
\end{align}

Moreover, as Appendix~\ref{sec:appendix} shows, the moments defined
in Eq.~(\ref{eq:81}) are related to the ones in Eq.~(\ref{eq:79}):
\begin{align}
  \label{eq:83}
  \dfrac{\pi^2}2\mclu{0}{11}(T) = \dfrac{2}{\rho h} - \mclu{0}{00}(T), 
\end{align}
and
\begin{align}
  \label{eq:84}
  \dfrac{\pi^2}2\mclu{2}{11}(T) = \dfrac{2\pi^2}{3\rho h} - \mclu{2}{00}(T).
\end{align}

Only the moments $\mclu{0}{00}(T)$, $\mclu{1}{01}(T)$, and $\mclu{2}{00}(T)$ need
be computed, therefore, to determine the right-hand side of
Eq.~(\ref{eq:78}). For $j=0$, given that $\mclu{0}{01}=0$, the
equality is equivalent to the expression
\begin{align}
  \label{eq:85}
  \mcl{0}(T) = (\alpha_0^2-{\tilde\alpha}_1^2)\mclu{0}{00}(T) 
  + \dfrac{2}{\rho h}{\tilde\alpha}_1^2, 
\end{align}
with the shorthand ${\tilde\alpha}\w1\equiv \sqrt{2}\alpha\w1/\pi$.

For $j=1$, Eq.~(\ref{eq:78}) amounts to
\begin{align}
  \label{eq:86}
  \mcl{1}(T) = 2\alpha\w{0}{\tilde\alpha}\w{1}\mclu{1}{01}(T),
\end{align}
and for $j=2$, to
\begin{align}
  \label{eq:87}
  \mcl{2}(T) = (\alpha_0^2-{\tilde\alpha}_1^2)\mclu{2}{00}(T) 
  + \dfrac{2\pi^2}{3\rho h}{\tilde\alpha}_1^2.
\end{align}

As already explained, the matrix elements $\mel{m}{\phi\w{k}}{n}$ ($k=0,1$) and the eigenvalues $E\w{m}$ and $E\w{n}$ on the right-hand sides of Eqs.~(\ref{eq:79}), (\ref{eq:80}), and (\ref{eq:81}) are universal functions of the energy scaled by $\kb\tk$. The three moments $\mclu{0}{00}$, $\mclu{1}{01}$, and $\mclu{2}{00}$ are universal functions of the ratio $T/\tk$. Equations~(\ref{eq:85})-(\ref{eq:87}) map the energy moments $\mcl{0}$, $\mcl{1}$, and $\mcl{2}$ onto $\mclu{0}{00}$, $\mclu{1}{01}$, and $\mclu{2}{00}$, respectively . The following analysis shows the linear coefficients $\alpha\w{0}$ and $\tilde{\alpha}\w{1}$ to be trigonometric functions of the fixed-point phase shifts.

\subsection{Linear coefficients} \label{sec:line-mapp-coeff}
At the LM and FL, the spectral densities for the operator $f\w{0}$ are \cite{2009SYO67006} \begin{align}
  \label{eq:88}
  \rho\w{0} = \rho\cos^2\bar\delta\qquad(\mbox{LM})\\
  \rho\w{0} = \rho\cos^2\delta\qquad(\mbox{FL}).
\end{align}

Equation~(\ref{eq:59}) relates the LM phase shift $\bar{\delta}$ to the FL phase shift $\delta$. Substitution of Eq.~(\ref{eq:88}) for the spectral density on the right-hand side of Eq.~(\ref{eq:63}) followed by integration yields the following limits for the lowest-order moment: \begin{align}\label{eq:89}
    \mcl{0} =
  \begin{cases}
    \dfrac{2}{h} \sin^2\delta&\qquad(\mbox{LM})\\[8pt]
    \dfrac{2}{h} \cos^2\delta&\qquad(\mbox{FL})
  \end{cases}.
\end{align}

The right-hand sides of Eq.~(\ref{eq:89}) can now be combined with Eq.~(\ref{eq:85}) to relate the high- and low-temperature limits of the universal moment $\mclu{0}{00}$ to the phase shift: \begin{align}
  \label{eq:90}
  \dfrac{2}{h} \sin^2\delta = (\alpha_0^2-{\tilde\alpha}_1^2)\mclu{0}{00}(LM)+
  \dfrac{2}{h}{\tilde\alpha}_1^2
  \qquad(\mbox{LM}), 
\end{align}
and
\begin{align}\label{eq:91}
  \dfrac{2}{h} \cos^2\delta = (\alpha_0^2-{\tilde\alpha}_1^2)\mclu{0}{00}(FL)+
  \dfrac{2}{h}{\tilde\alpha}_1^2
  \qquad(\mbox{FL}).
\end{align}

The universal moment $\mclu{0}{00}(T/\tk)$ is proportional to the SCD conductance for the symmetric model \cite{2009SYO67006} and hence drops from $\mclu{0}{00}(LM)=2/h$ at the LM to $\mclu{0}{00}(FL)=0$ at the FL. Equations~(\ref{eq:90})~and (\ref{eq:91}) therefore reduce to the equalities $\alpha_0^2=\sin^2\delta$, and ${\tilde\alpha}_1^2=\cos^2\delta$, respectively, which determine the absolute values of $\alpha\w{0}$ and ${\tilde\alpha}\w{1}$.

To determine the signs, we set $J=0$ on the right-hand side
of Eq.~\eqref{eq:69}. The resulting Hamiltonian is quadratic and can be diagonalized analytically \cite{YSO2009:235317}. It is then a simple matter to evaluate the matrix elements on both sides of Eq.~\eqref{eq:76}, from which it follows that that $\alpha\w{0}=\cos{\bar\delta}$ and
${\tilde\alpha}\w{1}=-\sin{\bar\delta}$. Equation~(\ref{eq:59}) then expresses the two
coefficients as trigonometric functions of the FL phase shift:
\begin{align} \label{eq:92}
  \alpha\w{0} = -\sin\delta,
\end{align}
and
\begin{align}\label{eq:93}
  {\tilde\alpha}\w{1} =-\cos\delta.
\end{align}

Substitution on the right-hand side of Eq.~(\ref{eq:86}) yields the
mapping between the energy moment $\mcl{1}$ and the universal
moment $\mclu{1}{01}$:
\begin{align}
  \label{eq:94}
  \mcl{1}(T/\tk) = \sin(2\delta)\mclu{1}{01}(T/\tk).
\end{align}

Likewise, substitution of Eqs.~(\ref{eq:92})~and (\ref{eq:93}) on the
right-hand sides of Eqs.~(\ref{eq:85})~and (\ref{eq:87}) determines
the coefficients mapping $\mcl{0}$ and $\mcl{2}$ onto the universal
moments $\mclu{0}{00}$ and $\mclu{2}{00}$,
\begin{align}
  \label{eq:95}
    \mcl{0}(T/\tk) = -\cos(2\delta)\mclu{0}{00}(T/\tk)+ \dfrac{1}{h}(1+\cos2\delta),
\end{align}
and
\begin{align}
  \label{eq:96}
      \mcl{2}(T/\tk) = -\cos(2\delta)\mclu{2}{00}(T/\tk)+ \dfrac{\pi^2}{3h}(1+\cos2\delta),
\end{align}
respectively.

The phase shift for the symmetric Hamiltonian is
$\delta=\pi/2$. Equations~(\ref{eq:94})-(\ref{eq:96}) then reduce
to $\mcl{1}(T/\tk)=0$ and $\mcl{j}(T/\tk) = \mclu{j}{00}(T/\tk)$ ($j=0,2)$, as expected.

The three expressions map the three energy moments onto universal functions. They reduce all temperature dependence to universal functions, which we need to compute only once. At the symmetric point, $\mcl{1}$ vanishes, while $\mcl{0}$ and $\mcl{2}$ reduce to the universal functions. Particle-hole asymmetry makes $\mcl{1}$ nonzero and flattens the temperature dependence of the other two moments. In all cases, the mapping is linear, with slopes and intercepts that depend on the ground-state phase shift only. The following section exhibits eNRG data confirming these findings.
%section 2 
%%%%%%%%%%%%%%%%%

%section 3
\section{Numerical results}\label{sec:results}
 
Table~\ref{tab:RSNRGRUNS} lists the gate potentials defining three eNRG runs with fixed Coulomb repulsion $U=10\,t$ and level width $\Gamma=0.40\,t$. The ratio $\Gamma/\min(|V\w{g}|, U+V\w{g})$ is $0$, $0.11$, and $0.2$ in runs $A$, $B$, and $C$, respectively. Runs $A$ and $B$ lie well within the Kondo regime, a condition that warrants the mappings to the universal functions and enhances the departures from the Wiedemann-Franz law, as discussed in Sec.~\ref{sec:wiedmann}. By contrast, the proximity of run $C$ to the charge-degeneracy point $V\w{g}=-U$ gives rise to significant deviations from universality.

\begin{table}
\caption{\label{tab:RSNRGRUNS} Gate potential, phase shifts and Kondo temperatures for the eNRG runs discussed in the text.}
\begin{ruledtabular}
\begin{tabular}{rcllll}
Run & Symbol & $V\w{g}/t$ & $\delta/\pi$  & $k\w{B} \tk/t$ & $\rho J$\\ \hline
A &\begin{tikzpicture} \draw[ultra thick, color1] (0,0)to(0.5,0); \end{tikzpicture} & -5.0 & 0.500 & $\num{6.4e-5}$&0.65\\%J=0.116 , from T_K
B& \begin{tikzpicture} \draw[ultra thick, densely dotted, color2] (0,0)to(0.5,0); \end{tikzpicture} & -6.5 & -0.491 & $\num{1.5e-4}$&0.71\\% J = 0.13
  C& \begin{tikzpicture} \draw[ultra thick, densely dashed, color3]  (0,0)to(0.5,0);
  \end{tikzpicture} & -8.0 & -0.470 & $\num{2.0e-3}$&1.01 % J = 0.18
\end{tabular}
\end{ruledtabular}
\end{table}

%\twocolumngrid
\subsection{Thermoelectric properties} \label{sec:thermoelectricproperties}

\begin{figure}
  \includegraphics[width = 0.8\linewidth]{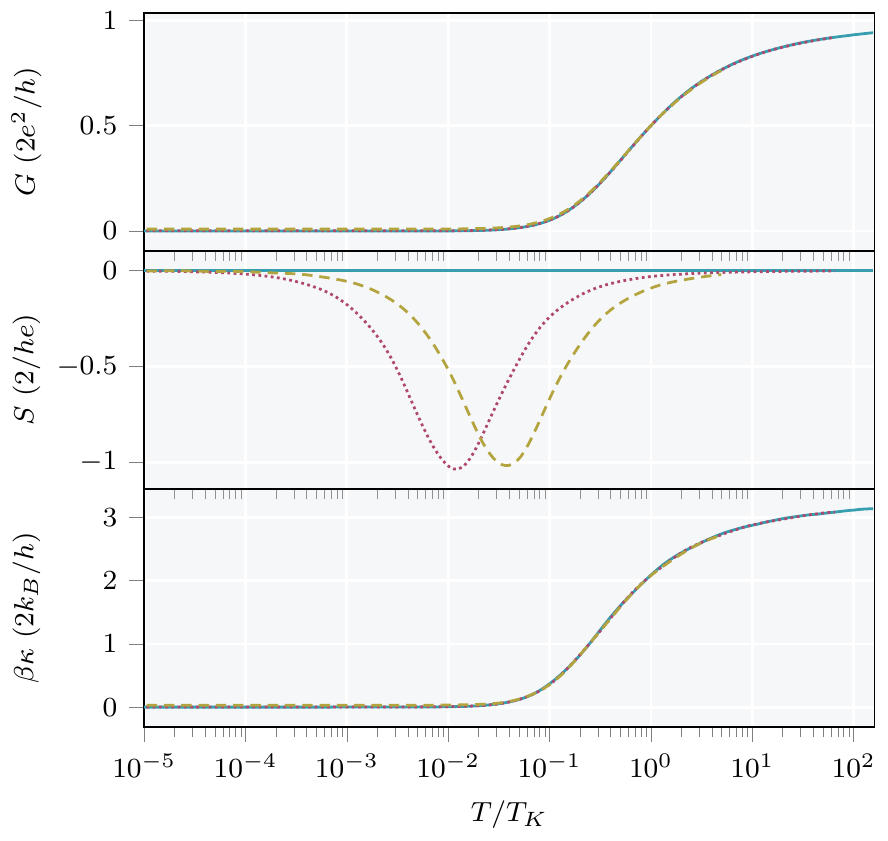}
  \caption[gsk]{Temperature dependence of the conductance (top), thermopower (central), and thermal conductance divided by the temperature (bottom panel), computed from Eqs.~\eqref{eq:66}, \eqref{eq:67},~and \eqref{eq:68}, respectively. In each panel, the solid cyan, dotted magenta, and dashed olive lines represent runs $A$, $B$, and $C$, respectively.}
  \label{thermoelectric}
\end{figure}

Along with the gate potential $V\w{g}$, Table~\ref{tab:RSNRGRUNS} presents the Kondo coupling $J$ computed from Eq.~\ref{eq:70}, the Kondo thermal energy $\kb \tk$, the ground-state phase shift $\delta$, and the style of the line representing each run in Figs.~\ref{thermoelectric}, and \ref{universalcurvesL0}-\ref{WF_law}. Figure~\ref{thermoelectric} displays the numerically computed conductance, thermopower, and thermal conductance as functions of temperature scaled by the Kondo temperature.

The electrical-conductance $G(T/\tk)$ and the thermal-conductance $\beta \kappa(T/\tk)$ curves are similar. As the temperature drops, both functions decay monotonically to zero, from the high-temperature plateaus of $G(T\gg \tk)=\gtwo$ and $\beta\kappa(T/\tk)=\pi^2/3$. Physically, at high temperatures, the conduction electrons flow ballistically across the quantum wire; the antiferromagnetic interaction with the dot magnetic moment offers little resistance to transport. Upon cooling, the Kondo cloud is gradually formed, and the progressively stronger coupling between conduction states and the dot orbital obstructs conduction through the central region of the wire.

The three conductance curves in the top panel of Fig.~\ref{thermoelectric} are nearly undistinguishable, a coincidence that turns our attention to the phase shifts in Table~\ref{tab:RSNRGRUNS}. The tabulated phase shifts are close to $\pi/2$ because the three runs are in the Kondo regime: on the scale of $\Gamma$, runs $A$ and $B$ are far from the charge-degeneracy point $V\w{g}=-U$, while run $C$ is moderately distant from it.

How does that affect the conductance? Equation.~\eqref{eq:79}  shows that the moment $\mcl{0}(T)$ and, hence, the conductance $G(T)$ are parametrized by $\cos{2\delta}$. In the Kondo regime, this trigonometric function lies close to its minimum and is, hence, nearly independent of $\delta$. The three curves in the top panel are, therefore, practically congruent.

For the same reason, the three curves in the bottom panel are virtually identical. 
In the middle panel, however, the distinctions are patent. The Seebeck coefficient monitors the difference between electron and hole conduction. Unlike the conductances, the thermopower changes sign under the particle-hole transformation. $S(T)$ vanishes in run $A$, for the symmetric model, with $V\w{g}+U/2=0$, depicted by the solid cyan line in the figure. For $V\w{g} + U/2 <0$, as in runs $B$ and $C$, the thermopower is negative. For $V\w{g}+ U/2>0$ (not shown), it is positive. 

Physically, the thermoelectric effect stems from transport across the quantum wire assisted by virtual excitations to the quantum dot. In the Kondo regime, the dot occupation is close to $n\w{d}=1$, as Fig. ~\ref{QD_config} indicates.  With $V\w{g}+U/2<0$ ($V\w{g}+U/2>0$) the dominant excitation is a transition from one of the central columns to the rightmost (leftmost) one, which transfers an electron (a hole) to the quantum dot; the resulting Seebeck coefficient is negative (positive).

At high temperatures, independently of the sign of $V\w{g}+U/2$, electrons flow freely across the wire. Nonetheless, the weak coupling to the dot reduces $S(T\gg\tk)$ to zero. At low temperatures, the dot is strongly coupled to the wire, but the Kondo cloud blocks transport. Only at intermediate temperatures can the Seebeck coefficient differ significantly from zero.

The thermopower is the ratio on the right-hand side of Eq.~\eqref{eq:67}, between the moments $\mcl{1}(T)$ and $\mcl{0}(T)$. If the ratio were proportional to a universal function, only the amplitude of the plot would depend on $V\w{g}$. The numerator $\mcl{1}(T)$ is, in fact, proportional to the universal moment $\mclu{1}{01}(T/\tk)$, but the denominator $\mcl{0}(T)$ is neither universal nor proportional to a universal function, as Eq.~\eqref{eq:95} shows.    The weak dependence of the denominator on the phase shift $\delta$ is sufficient to shift the symmetric maximum from $0.01\tk$ to approximately $0.04\tk$ as the phase shift is reduced from $\delta=0.49\pi$ to $\delta=0.47\pi$.

\subsection{Universal moments} \label{sec:universalresults}
\begin{figure}
  \includegraphics[width = 0.8\linewidth]{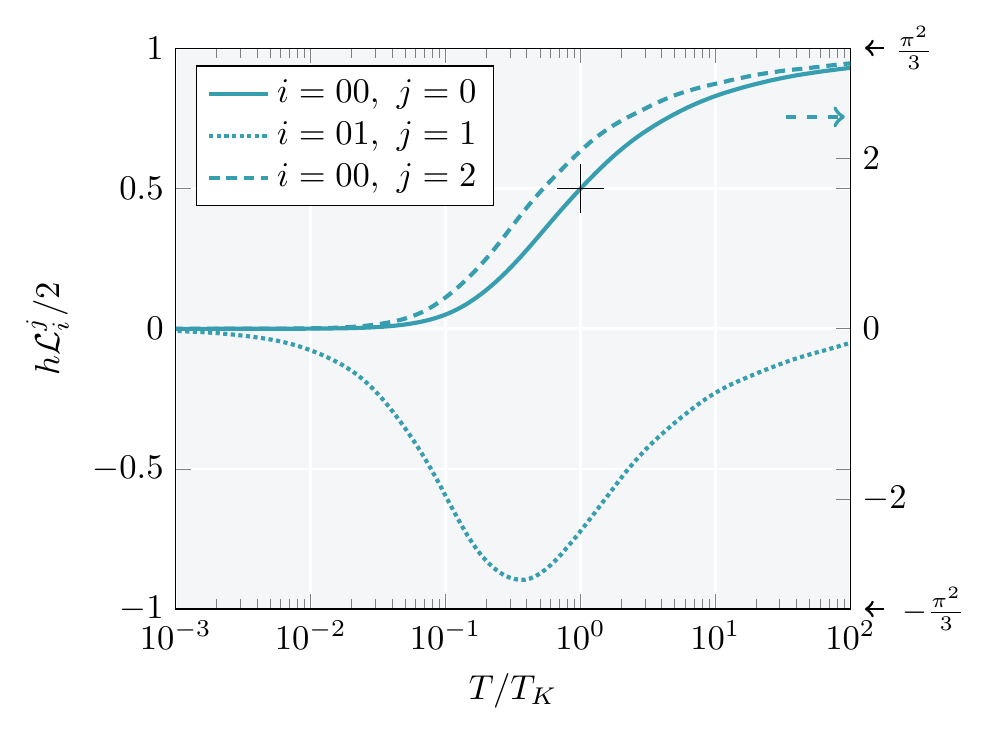}
  \caption[universal]{Universal energy moments $\mclu{0}{00}$, $\mclu{1}{01}$, and $\mclu{2}{00}$ as functions of the temperature scaled by $\tk$. The hairlines mark the definition $\mclu{0}{00}(T/\tk=1) = 1/h$ of the Kondo temperature.}
  \label{universalcurves}
\end{figure}

Figure~\ref{universalcurves} shows the thermal dependence of the universal energy moments $\mclu{0}{00}$, $\mclu{1}{01}$, and $\mclu{2}{00}$ onto which the temperature dependences of the transport moments $\mcl{0}$, $\mcl{1}$, and $\mcl{2}$ are linearly mapped, respectively. The solid line depicts $\mclu{0}{00}(T/\tk)$, which is proportional to the conductance at the symmetric point. As the temperature drops, $\mclu{0}{00}$ diminishes monotonically to zero, from the ballistic limit $\mclu{0}{00}(T\gg\tk)=2/h$. The hairlines identify the halfway point $\mclu{0}{00}(T=\tk) \equiv 1/h$, which defines the Kondo temperature.

The dashed line depicts the analogous decline of the universal moment $\mclu{2}{00}(T/\tk)$. The curve decays to zero from the high-temperature plateau $\mclu{2}{00}=2\pi ^{2}/(3h)$, and crosses its half-maximum $\pi^2/(3h)$  at $T\approx \tk/2$.

The temperature dependence of the universal moment $\mclu{1}{01}(T/\tk)$ is conspicuously distinct. The moment vanishes at high and low temperatures, and becomes negative  throughout the crossover from the LM to the FL. The dotted line in Fig.~\ref{universalcurves} displays a broad, nearly symmetric minimum centered at $T\approx 0.4\tk$.

\begin{figure}
  \includegraphics[width = 0.8\linewidth]{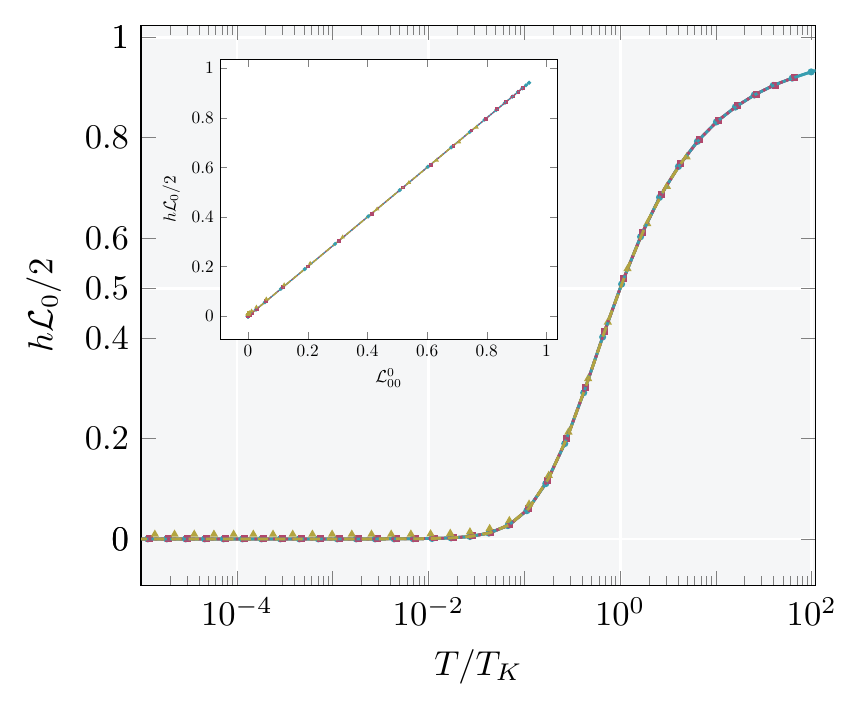}
  \caption[L0]{First energy moment as a function of the temperature scaled by $\tk$. The filled circles, squares, and triangles, fitted by the solid cyan, dotted magenta, and dashed olive lines depict the numerical data resulting from eNRG runs $A$, $B$, and $C$, respectively. The lines represent Eq.~\eqref{eq:95} for the tabulated phase shifts in the three runs . The inset shows the energetic moment as a function of the universal curve $\mclu{0}{00}(T/\tk)$ for the three runs. The excellent agreement between the circles and the straight line fitting them witnesses the linearity of the mapping.}
  \label{universalcurvesL0}
\end{figure}

Figure~\ref{universalcurvesL0} compares numerically computed moments with the mapping~\eqref{eq:95}. The moments $\mcl{0}(T)$ computed in the three runs in Table~\ref{tab:RSNRGRUNS} are represented by circles, plotted as functions of the ratio $T/\kb\tk$ in the range $\num{1e-5}<T/\tk<\num{1e2}$. The solid lines show the right-hand side of Eq.~\eqref{eq:95}, parametrized by the tabulated $\delta$. The phase shifts being close to each other, the three curves are nearly coincident. Even the small differences between moments correspondent to phase shifts only a few percent apart are accurately reproduced by the universal mappings, however.

For better comparison, the inset of Fig.~\ref{universalcurvesL0} plots the computed moments as functions of the universal moment $\mclu{0}{00}$. The excellent agreement with the straight lines representing Eq.~\eqref{eq:95} for the pertinent phase shifts attests the accuracy of the data. The same procedure can be applied to experimental results, as illustrated by analyses focused on data collected in side-coupled devices \cite{2009SYO67006} or single-electron transistors \cite{2018ZaO136}. More on that in Sec.~\ref{sec:conductance}.

\begin{figure}
  \includegraphics[width = 0.8\linewidth]{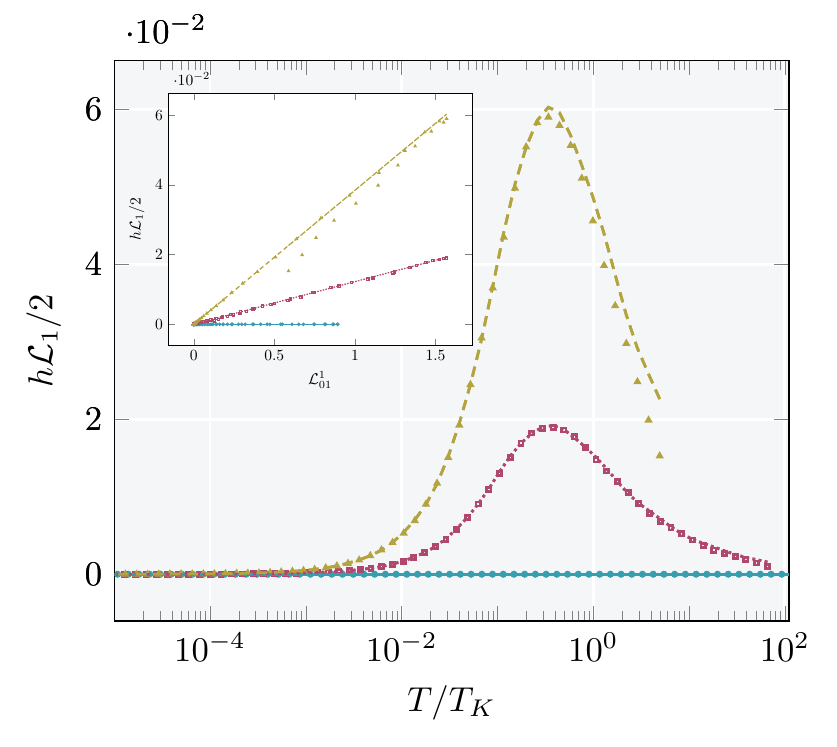}
  \caption[L1]{Temperature dependence of the second energy moment. The plots follow the symbol convention in Fig.~\ref{universalcurvesL0}. The inset shows $\mcl{1}(T/\tk)$ as a function of the universal moment $\mclu{1}{01}(T/\tk)$, to probe Eq.~\eqref{eq:94} visually. Both the main plot and inset display distinct deviations from universal behavior at high temperatures, an indication that the relatively large $|V\w{g}+U/2|$ has pushed run $C$ too close to the charge degeneracy point $V\w{g}=-U$.}
  \label{universalcurvesL1}
\end{figure}

Figure~\ref{universalcurvesL1} shows the analogous plots for the $\mcl{1}(T)$ moment. The results from run~$A$ are shown for completeness only, because $\mcl{1}(T)$ vanishes at all temperatures. The moments from runs~$B$~and $C$ are positive because $V\w{g}+U/2<0$, which makes the Schrieffer-Wolff scattering potential $\bar{W}$ negative. The LM phase shift $\bar\delta$ is hence positive, the FL phase shift $\delta$ is negative, and so is the factor multiplying $\mclu{1}{01}$ on the right-hand side of Eq.~\eqref{eq:94}. For $V\w{g}+U/2>0$ (not shown), the moment is negative at all temperatures, a reminder that the thermopower is very sensitive to particle-hole asymmetry.

The dotted magenta lines in the main plot and inset show good agreement with the filled magenta squares. The small deviations at the highest temperatures are contributions from the $\order{\rho\kb T}$ terms neglected in the derivation of Eq.~\eqref{eq:94}, which become significant for $\kb T\agt\num{5e-3}$, that is, for $T\agt 5\tk$ in run~$B$. Manifest deviations with the same origin separate the olive triangles representing the moments computed in run~$C$ from the olive dashed line. Since run $C$ is relatively close to the charge-degeneracy point $V\w{g}=-U$, the Kondo temperature $\kb\tk = \num{2e-3}t$ is fairly high, and the discrepancies become visible even below the Kondo temperature.

\begin{figure}
  \includegraphics[width = 0.8\linewidth]{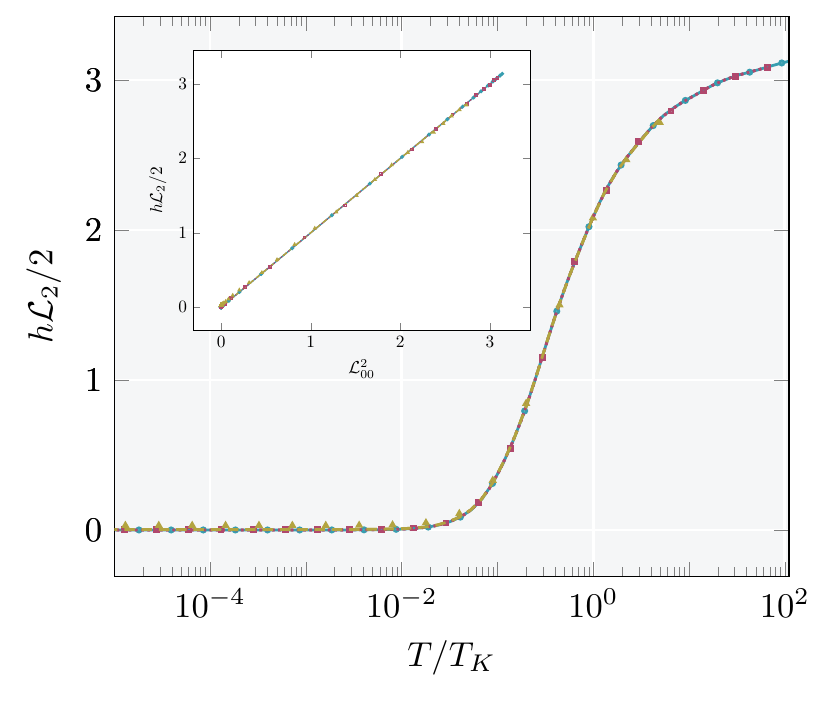}
  \caption[L2]{Temperature dependence of the third energy moment. The plots follow the symbol convention in Figs.~\ref{universalcurvesL0}~and \ref{universalcurvesL1}. The inset shows $\mcl{2}(T/\tk)$ as a function of the universal moment $\mclu{2}{00}(T/\tk)$ to corroborate Eq.~\eqref{eq:96}.}
  \label{universalcurvesL2}
\end{figure}

The filled circles, squares, and triangles in the main plot of Fig.~\ref{universalcurvesL2} show the $\mcl{2}$ computed in runs~$A$, $B$, and $C$ as functions of the temperature scaled by the Kondo temperature, respectively. The solid, dotted, and dashed lines represent the right hand side of Eq.~\eqref{eq:95} with $j=2$ and the phase shifts listed in Table~\ref{tab:RSNRGRUNS} for runs $A$, $B$, and $C$, respectively. The inset shows the same moments and universal mappings as functions of the universal moment $\mclu{2}{00}(T/\tk)$. All curves, including those in the inset, are in close analogy with the plots in Fig.~\ref{universalcurvesL0}. In particular, in contrast with the main plot in Fig.~\ref{universalcurvesL1}, the three curves are close to congruence, because the right-hand side of Eq.~\eqref{eq:95} is insensitive to changes in the phase shift near $\delta=\pi/2$. Another contrast with Fig.~\ref{universalcurvesL1} is the excellent agreement between each line and the corresponding set of circles, squares, or triangles, which indicates that the terms of $\order{\rho\epsilon}$ dropped in the derivation of Eq.~\eqref{eq:95} make smaller contributions to $\mcl{0}$ and $\mcl{2}$ than to $\mcl{1}$.

\subsection{Wiedemann-Franz law} \label{sec:wiedmann}
The Wiedemann-Franz law states that the ratio between the thermal and electrical conductances is proportional to the temperature:
\begin{align}
  \label{eq:97}
  \dfrac{\kappa(T)}{G(T)} = L\w{0} T,
\end{align}
where $L\w{0} \equiv (\pi^{2}/3) (k\w{B}/e)^2$ denotes the Lorenz ratio \cite{AsM1976solid}.

This expression of the equivalence between energy and charge transport results from rigorous expressions for the electrical and thermal conductances of free electrons. It is, therefore, reliable at Fermi-liquid fixed points. Here, the law is valid at the fixed points of the renormalization-group transformation $\tau^2$.  At the LM and FL, Eq.~(\ref{eq:97}) follows from Eqs.~\eqref{eq:95}~and \eqref{eq:96}, which read
\begin{subequations}\label{eq:98}
  \begin{align}
    \mcl{0} &= \dfrac{2}{h}\sin^2\delta&\ncr
            &&\qquad(LM)\ncr
    \mcl{2} & =\dfrac{2\pi^2}{3h}\sin^2\delta&\nonumber
  \end{align}
\end{subequations}
and
\begin{subequations}\label{eq:99}
  \begin{align}
    \mcl{0} &= \dfrac{2}{h}\cos^2\delta&\ncr
            &&\qquad(FL)\ncr
    \mcl{2} & =\dfrac{2\pi^2}{3h}\cos^2\delta&\nonumber
  \end{align}
\end{subequations}

\begin{figure}
    \includegraphics[width = 0.8\linewidth]{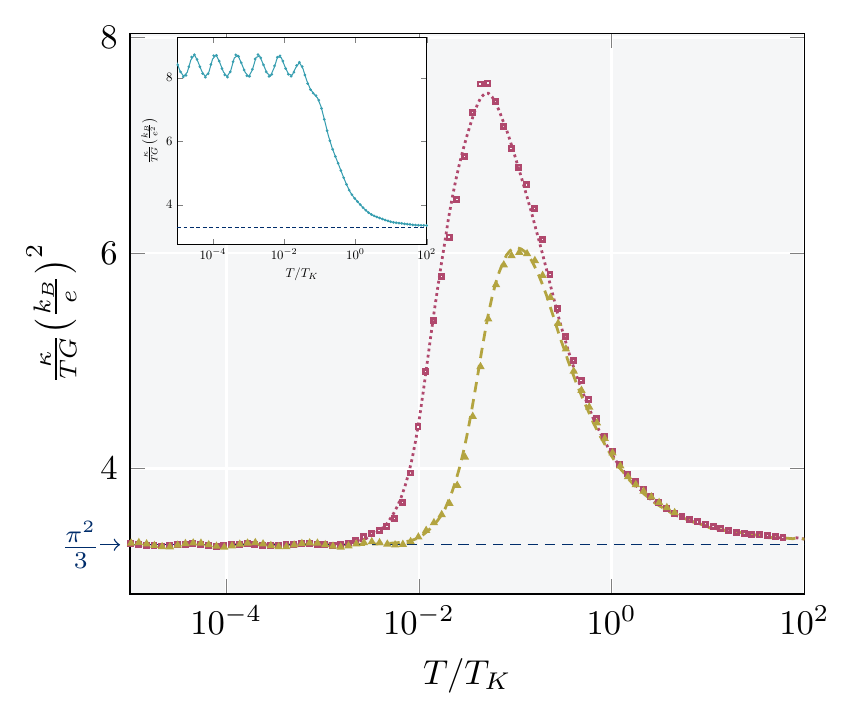}
    \caption[WF]{Wiedemann-Franz ratio as a function of the temperature scaled by $\tk$ for runs $B$ and $C$. The symbol convention follows Fig.~\ref{thermoelectric}. 
      At high and low temperatures the plots approach the Lorenz number $L\w{0} = (\pi^{2}/3)(k\w{B}/e)^2$. The inset shows the results from run $A$, which are markedly different for $T<\tk$ because both $G$ and $\beta\kappa$ vanish as $T\to0$ at the symmetric point. }
    \label{WF_law}
\end{figure}

As one might expect, Fig.~\ref{WF_law} shows that the proportionality between $G(T)$ and $\beta\kappa(T)$ breaks down in the intermediate temperature range. The ratio $\beta\kappa/G$ peaks below the Kondo temperature, an indication that, as the temperature is reduced past $\tk$, the Kondo cloud starts to block charge transport more efficiently than to obstruct energy transport; at lower temperatures, the two forms of obstruction become comparable.

The peak is less pronounced for $V\w{g}=-8U$ than for $V\w{g}=-6.5U$. This is expected from Eqs.~\eqref{eq:95}~and \eqref{eq:96}, which show that, as functions of the phase shift, the differences $\Delta G$ and $\Delta\kappa$, between the high- and low-temperature electrical and thermal conductances are maximized at the symmetric point, with $\delta=\pi/2$. As $|V\w{g}+U/2|$ grows and $|\delta|$ is reduced, the differences shrink, and the thermal dependences of the conductances become flatter. If the phase shift were $\delta=\pi/4$, both conductances would be independent of $T$, and Eq.~(\ref{eq:97}) would be valid at all temperatures. 

At the symmetric point $V\w{g}=-U/2$, particle-hole symmetry forces the phase shift to be $\pi/2$. In the Kondo regime, as the numbers in Table~\ref{tab:RSNRGRUNS} shows, the phase shift stays close to $\pi/2$. Values closer to $\pi/4$ can only be found in the vicinity of the charge-degeneracy condition $V\w{g}=-U$, where Eqs.~\eqref{eq:95}~and \eqref{eq:96} are invalid, because universality breaks down.

Nonetheless, the trend to thermal independence, of which $\delta=\pi/4$ is the extreme, emerges in the Kondo regime as the gate potential grows away from particle-hole symmetry. As a result, the deviations from Wiedemann-Franz behavior become less pronounced, and the peak drawn by the olive triangles and dashed line in Fig.~\ref{WF_law} is substantially smaller than the one drawn by the magenta squares and dotted line, even though the phase shifts are by no means close to $\pi/4$.

The inset of Fig.~\ref{WF_law} shows the Wiedemann-Franz ratio as a function of temperature for run $A$.  With $\delta=\pi/2$, both $G$ and $\beta\kappa$ vanish at $T=0$s. The ratio between the two conductances is hence determined by their low-temperature expansions and deviates from the Lorenz number. The oscillations in the plot are artifacts of the discretization that have been only partially eliminated by the averaging procedure.

\section{Comparison with experiments}
\label{sec:comp-with-exper}

The central results in Sec.~\ref{sec::universality}, Eqs.~\eqref{eq:94}-\eqref{eq:96}, aid the interpretation of measurements. Accurate computation of the universal functions $\mclu{0}{00}(T/\tk)$, $\mclu{1}{01}(T/\tk)$, and $\mclu{2}{00}(T/\tk)$ requires small computational effort. Once the three universal functions have been calculated, it becomes possible to fit the temperature dependence of transport properties measured in Kondo systems, with two adjustable parameters: the Kondo temperature and phase shift. With more than one parameter, the linearity of the mappings~\eqref{eq:94}-\eqref{eq:96} acquires special significance, because it supports algorithms that expedite the fitting.

\subsection{Conductance}
\label{sec:conductance}

Since the conductance is proportional to $\mcl{0}$, it follows from Eq.~\eqref{eq:95} that plots of $G(T/\tk)$ as functions of $\mclu{0}{00}(T/\tk)$ are straight lines. In practice, $\tk$ is unknown. However, given that the curvature of the plot reverses trial values for the Kondo temperature grow past $\tk$, a bisection algorithm readily yields the Kondo temperature. The slope or the intersection of the straight line then determines the phase shift. A number of examples dealing with conductance data from SET or side-coupled devices have been presented~\cite{2009SYO67006, 2018ZaO136,2021ARS085112}. The algorithm yields excellent fits and allows accurate determination of the Kondo temperature, even though background currents of unknown origin increment the SCD conductances, while contact asymmetries restrict the SET conductances to maxima substantially below the conductance quantum.

The interpretation of conductance curves is relatively simple, given the abundance of data of remarkable quality, collected in finely engineered devices. The literature focused on the other transport properties is scantier and less diverse.

\subsection{Thermopower}
\label{sec:thermopower}

In contrast with the conductance, the thermopower maps nonlinearly onto the universal moments $\mclu{0}{00}$ and $\mclu{1}{01}$. 
A linear relation can, nevertheless, be established \cite{2021ARS085112}. To this end,  we substitute the right-hand side of Eq.~\eqref{eq:94} for $\mcl{1}$ and the right-hand side of Eq.~\eqref{eq:95} for $\mcl{0}$ on the right-hand side of Eq.~\eqref{eq:67}. Straightforward manipulations then show that
\begin{align}
  \label{eq:100}
  \dfrac{h\mclu{1}{01}(T/\tk)}{S(T/\tk)} = \cot(\delta)-\cot(2\delta)\dfrac{h}2\mclu{0}{00}(T/\tk). 
\end{align}

The universal functions $\mclu{0}{00}(T/\tk)$ and $\mclu{1}{01}(T/\tk)$ are easily computed. Thus, given (i) a tabulation of thermopowers measured in a range of temperatures, and (ii) a trial Kondo temperature $\tk$, the left hand-side of Eq.~(\ref{eq:100}) can be computed and plotted as a function of the right-hand side. As in Sec.~\ref{sec:conductance}, bisection indexed by the curvature of the plot then determines the Kondo temperature and, subsequently, the phase shift. 

An example to demonstrate the effectiveness of this procedure seems warranted. No measurements of the thermopower in the side-coupled geometry have been reported, however. While numerous studies focused on the conductance of nanostructured devices are found in the literature from the last two decades, the other transport properties have received virtually no attention, and the two recent exceptions preferred the bridge geometry \cite{2018SJB206801,2019DMCp506}.

Substitutional alloys with low concentration of a magnetic species offer an attractive alternative, since Eqs.~\eqref{eq:66}, \eqref{eq:67},~and \eqref{eq:68} describe their transport properties in the dilute limit, up to a system dependent proportionality factor. Here, the discussion is centered on the rare-earth compound Lu$\w{0.9}$Yb$\w{0.1}$Rh$\w{2}$Si$\w{2}$ and the thermal dependence of its thermopower at moderately low temperatures \cite{2007Koh}. The crystal field of the lattice splits the ground-state multiplet of the free Yb ion into four doublets. The lowest doublet lies \SI{210}{K} below the first excited one. To reduce the contribution from the excited doublets, the following analysis will be restricted to temperatures below $\SI{75}{K}$. Under these conditions, the Yb becomes approximately equivalent to a spin-$1/2$ impurity coupled to the conduction electrons, a Kondo system, that is.

\begin{figure}[!th]
  \includegraphics[width=0.8\linewidth]{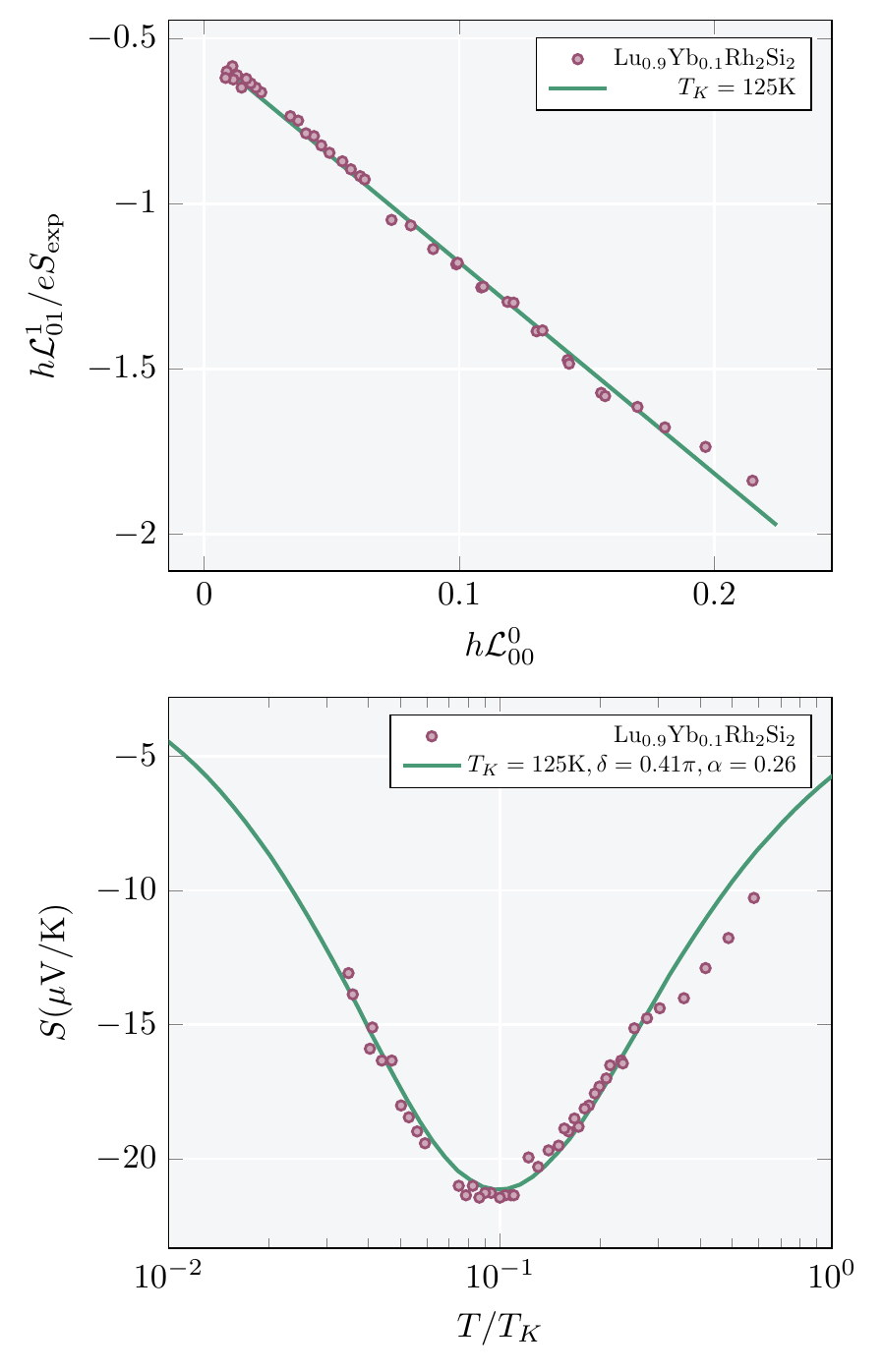}
  \caption[LuYbRhSi]{Thermopower of the Kondo system Lu$\w{0.9}$Yb$\w{0.1}$Rh$\w{2}$Si$\w{2}$. The filled circles are measurements reported in Ref.~\cite{2007Koh}, and the solid lines are optimized fits resulting from Eq.~\eqref{eq:101} with the constants in the legend of panel (b). Panel (a) plots the left-hand side of Eq.~\eqref{eq:101} as a function of $\mclu{0}{00}(T/\tk)$ to display the linear regression from which $\delta$ and $\alpha$ were obtained. Panel (b) shows the Seebeck coefficient as a function of temperature. }
  \label{fig:yb}
\end{figure}

Figure~\ref{fig:yb} depicts the thermal dependence of the thermopower. The circles represent the lab data \cite{2007Koh}, and the solid line is an optimized fit rooted in the mappings~\eqref{eq:94}~and \eqref{eq:95}. To account for the distinction between the one-dimensional side-coupled device and the lattice system, the left-hand side of Eq.~(\ref{eq:101}) was multiplied by an adjustable dimensionless parameter $\alpha$. This led to the expression
\begin{align}\label{eq:101}
  S(T) = \alpha \dfrac{2h}{e}\dfrac{h\mclu{1}{01}(T)}{\cot(2\delta)\mclu{0}{00}-2\cot(\delta)}
  \qquad(\mbox{lattice}). 
\end{align}

The solid line in the figure represents Eq.~(\ref{eq:101}) with $\tk=\SI{125}{K}$, $\delta=0.41$, and $\alpha=0.26$. The Kondo temperature was determined by the aforementioned bisection procedure, which converged to the straight line fitting Eq.~(\ref{eq:101}) in Fig.~\ref{fig:yb}(a). The slope $m=-\alpha\cot(2\delta)$ and the intersection $c= \alpha\cot(\delta)$ of that line then determined the phase shift $\delta$ and coefficient $\alpha$, and yielded the fit in panel (b). The systematic deviations separating the rightmost circles from the solid lines in both panels reflect the contributions from the excited doublets of the Yb ions, which are frozen out as the temperature is reduced below \SI{40}{K}.

The agreement in both panels confirms K\"{o}hler's  interpretation \cite{2007Koh}, which associated the relatively high thermopower in the substitutional compound with the Kondo effect. The Kondo temperature in Fig.~\ref{fig:yb} and the value in Ref.~\cite{2007Koh} are different, by an order of magnitude, because the definitions are different. Here, $\tk$ is defined by the equality $G(\tk)= \gtwo/2$, while K\"ohler associated the Kondo temperature with the maximum of $|S(T)|$. 

\subsection{Thermal conductance}
\label{sec:thermal-conductance}

The first term on the right-hand side of Eq.~\eqref{eq:68} is the energy moment $\mcl{2}(T)$, which maps linearly onto the universal moment $\mclu{2}{00}$. Even though the second term breaks the linearity, it is a simple matter to restore it. The right-hand sides of Eqs.~\eqref{eq:94}, \eqref{eq:95},~and \eqref{eq:96} can be substituted for $\mcl{1}$, $\mcl{0}$, and $\mcl{2}$ to convert Eq.~\eqref{eq:68} into a mapping that is approximately linear:
\begin{align}
  \label{eq:102}
  \beta \kappa\Big(\frac{T}{\tk}\Big) +&\dfrac{\Big({\mclu{1}{01}}(\frac{T}{\tk})\Big)^2} {\cot(\delta)-h\cot(2\delta)\mclu{0}{00}(T)}
  =\ncr
     &\dfrac{\pi^2}{3h}\cot(\delta) -\cot(2\delta)\mclu{2}{00}(T).
\end{align}

Equation~(\ref{eq:102}) is less convenient than Eq.~(\ref{eq:100}), or (\ref{eq:101}), because its left-hand side depends on the phase shift. An iterative procedure is now required to determine $\tk$ and $\delta$. The second term on the left-hand side vanishes for $\delta=\pi/2$, and is small in the Kondo regime. In the first iteration, that term is neglected, the left-hand side becomes analogous to Eq.~(\ref{eq:101}), and the bisection procedure described in Secs.~\ref{sec:conductance}~and \ref{sec:thermopower} yields a first estimate: $\tk=T_K^1$ and $\delta=\delta^1$. This concludes the first iteration.

In the $n$-th iteration, substitution of $T_K^{n-1}$ and $\delta^{n-1}$ for $\tk$ and $\delta$ turns the second term on the right-hand side of Eq.~(\ref{eq:102}) into a known function $f(T/T_K^{n-1})$. Given a trial $\tk$, the sum $\beta\kappa\w{\mbox{exp}}(T/\tk)+ f((T/T_K^{n-1})$ can then be depicted as a function of $\mclu{2}{00}(T/\tk)$. This defines a bisection procedure indexed by the curvature of the plot, which determines the improved estimates $T_K^n$ and $\delta^n$, and closes iteration $n$.

Unfortunately, lack of pertinent experimental data precludes presentation of an example. Measurements of $\kappa(T)$ in lattice systems have been reported, but the contribution of the Kondo cloud cannot be extricated from the phonon and electron-phonon contributions to $\kappa(T)$ \cite{2015BTBp012004}. Measurements in side-coupled devices seem therefore necessary before Eq.~(\ref{eq:100}) can be put to the test.

\section{Summary}
\label{sec:summary}

This paper presents an alternative formulation of the NRG procedure, abbreviated eNRG because it is based on an exponentially-growing sequence of blocks in real space, instead of on a logarithmic sequence of intervals in momentum space. Projection of the conduction-band Hamiltonian upon the resulting basis yields the discretized form~\eqref{eq:49}, analogous to the codiagonal Hamiltonian generated by the logarithmic discretization \cite{FirstUniversality}. The codiagonal coefficients $t\w{n}$ and $\bar t\w{n}$ in the two series decay exponentially as $n$ grows, and the identification $\lambda^2\equiv \Lambda$ makes the sequence $t\w{n}$ ($n=0,1,\ldots$) asymptotically proportional to the sequence $\bar t\w{n}$ ($n=0,1,\ldots$). The proportionality breaks down for small $n$ because the two discretizations are applied to distinct dispersion relations: $\epsilon\w{k}= -2t\cos(k)$ and $\epsilon\w{k}= D(k-k\w{F})$ in the eNRG and NRG approaches, respectively.

Another distinction is the flexibility afforded by the second parameter in the eNRG discretization. The offset $\offs$ controls the phase of the oscillations artificially added to the computed thermal dependence of physical properties. Averaging over two subsequent offsets eliminates such oscillations and yield accurate approximations to the continuum limit. In addition, larger offsets describe the conduction energies near the band edges more reliably. The high-temperature congruence between the eNRG-computed conductances and the exact results for the noninteracting model in Fig.~\ref{fig:1} offers an illustration.

The numerically computed thermal properties for the interacting model, in Sec.~\ref{sec:results}, survey the accuracy of the eNRG procedure. The eigenvalues and eigenvectors resultant from the iterative diagonalization of the model Hamiltonian yield the energy moments $\mcl{j}$ ($j=0,1,2$) from which the electrical conductance $G(T)$, the Seebeck coefficient $S(T)$, and the thermal conductance $\kappa(T)$ are computed. As Figs.~ \ref{universalcurvesL0}-\ref{universalcurvesL2} show, the numerical results for the three energy moments agree very well with Eqs.~\eqref{eq:94}, \eqref{eq:95}, and \eqref{eq:96} in the Kondo regime, which map the thermal dependence of the moments onto the three universal functions $\mclu{1}{01}(T/\tk)$, $\mclu{0}{00}(T/\tk)$, and $\mclu{2}{00}(T/\tk)$, respectively.

The mappings onto the universal functions provide insight. To dwell on this point, Sec.~\ref{sec:wiedmann} explains why the deviations from the Wiedemann-Franz law shrink as the model parameters move away from particle-hole symmetry.  The final section, discusses the algorithms exploiting the linearity of the mappings to the universal functions to extract the Kondo temperature and phase shift from experimental data. Examples targeting electrical conductance data having been presented in previous publications, and the absence of experimental data in the side-coupled geometry barring application to the thermal conductance, the illustration in Fig.~\ref{fig:yb} is focused on a measurement of the thermopower.

With exception of Fig.~\ref{fig:1}, the above-described results of the eNRG method are linked to universality and could have been obtained via NRG treatment. The eNRG is, however, more than a simple derivation of the NRG Hamiltonian. A real-space formulation is fitter to describe nanostructures than one in momentum space, especially when both involve projections upon incomplete bases. The eNRG construction is expected to describe the RKKY interaction between the magnetic moments of two impurities or quantum dots better than the NRG approach \cite{1996SLOp275}, for instance. Additional work is planned to unravel the full potential of the method.
% section 3
%%%%%%%%%%%%%%%%%

%section ackn 
\acknowledgments
We are grateful to Luiz Henrique B.~Guessi for suggestions and very helpful discussions. 
This work has been supported by CAPES grant number 88887.475410/2020-00, CNPq grant number 312239/2018-1, and FAPESP grant number 2017/26215-4.
%section ackn 
%%%%%%%%%%%%%%%%%

\appendix 
%section app 
\section{Derivation of Eq.~\eqref{eq:40}}
\label{sec:coeff-relat-oper}
The first term on the right-hand side of Eq.~\eqref{eq:39} reads
\begin{align}
  \label{eq:103}
  H^{\mbox{diag}}_{f\lambda} \equiv -t\sum_{n=1}^{\infty}\sum_{j=1}^{\lambda^n}
                           (\alpha\w{n,j}\alpha_{n,j}^*+\mbox{c.~c.}) \fd{n}\fn{n}.
\end{align}
The superscript is a reminder that $H^{\mbox{diag}}_{f\lambda}$ only comprises the diagonal terms of $H\w{f\lambda}$.

Since $H^{\mbox{diag}}_{f\lambda}$ has no counterpart in the original Hamiltonian, Eq.~\eqref{eq:30}, we wish to choose the coefficients $\alpha\w{n,j}$ so that the factor within parentheses in the summand on the right-hand side be equal to zero, for $n=1,2,\ldots$ and $j=1,2,\ldots,\lambda^n$. In other words, the $\alpha\w{n,j}$ must satisfy the condition
\begin{align}
  \label{eq:104}
  \operatorname{Re}(\alpha\w{n,j}\alpha_{n,j+1}^*)=0, 
\end{align}
from which it follows that the phases $\alpha\w{n,j}$ and $\alpha\w{n, j+1}$ differ by an odd multiple of $\pi/2$:
\begin{align}
  \label{eq:105}
    \phi\w{n, j+1} =\phi\w{n,j} + (2p+1)\dfrac{\pi}2,
\end{align}
where $p$ is an arbitrary integer.

The simplest expression satisfying Eq.~\eqref{eq:105} is
\begin{align}
  \label{eq:106}
  \phi\w{n,j} = \phi\w{n}+j\dfrac{\pi}2,
\end{align}
where $\phi\w{n}$ denotes a phase that is uniform within each cell $\mathcal{C}\w{n}$ ($n=1, 2, \ldots$).

The $\phi\w{n}$ remain to be fixed. To this end, let us consider the second term on the right-hand side of Eq.~\eqref{eq:39}:
\begin{align}
  \label{eq:107}
 H_{f\lambda}^{\mathrm{off}}= -t\sum_{n=0}^{\infty}(\alpha_{n,\lambda\y{n}}^*\alpha\w{n+1,1}\,\fd{n}\fn{n+1}+\hc),
\end{align}
where the superscript is a reminder that $H_{f\lambda}^{\mathrm{off}}$ only comprises the off-diagonal terms of $H\w{f\lambda}$.

The phase of the factor multiplying $\fd{n}\fn{n+1}$ in the summand on the right-hand side of Eq.~\eqref{eq:107} is
\begin{align}  \label{eq:108}
  \phi\w{n+1, 1}-\phi\w{n, \lambda\y{n}} = \phi\w{n+1} -\phi\w{n} + \dfrac{\pi}2(1-\lambda\y{n}), 
\end{align}
which, in view of Eq.~\eqref{eq:34}, can be written in the form
\begin{align}
  \label{eq:109}
  \phi\w{n+1, 1}-\phi\w{n, \lambda\y{n}} = \phi\w{n+1} -\phi\w{n} + \dfrac{\pi}2(1-\gb{n+1}+\gb{n}).
\end{align}

We want to make the coefficient of $\fd{n}\fn{n+1}$ on the right-hand side of Eq.~\eqref{eq:107} real positive. The phase difference~\eqref{eq:109} must therefore be an odd multiple of $\pi$. The simplest expression satisfying this condition is Eq.~\eqref{eq:40}.

\section{Derivation of Eqs.~\eqref{eq:83}~and \eqref{eq:84}}
\label{sec:appendix} 
Equations~\eqref{eq:83}~and \eqref{eq:84} are immediate consequences
of a linear relation between the spectral densities of the operators $\phi\w{0}$
and $\phi\w{1}$. To establish this relation, we consider the symmetric
Anderson Hamiltonian and write down the Dyson equation for
the retarded conduction-electron Green's function:
\begin{equation}\label{eq:110}
    \mathbb{G}_{kk'}^{S}(\epsilon) =
    \mathbb{G}_{k}^{0}(\epsilon)\delta_{kk'}
    + \frac{V^{2}}{N} \mathbb{G}^{0}_{k}(\epsilon)
    \mathbb{G}^{S}_{d}(\epsilon) \mathbb{G}^{0}_{k'}(\epsilon),
  \end{equation}
  where  the superindex $S$ indicates association with the symmetric model, $\mathbb{G}^{S}_{d}(\epsilon)$ is the retarded dot-level Green's function, and $\mathbb{G}_k^0$ is the free conduction-electron Green's function:
  \begin{equation}\label{eq:111}
    \mathbb{G}_{k}^{0} = \frac{1}{\epsilon - \epsilon_{k} + i\eta}
\end{equation}

We want to determine the spectral densities of the operators $\phi\w{0}$ and $\phi\w{1}$, which are linear combinations of the LM eigenoperators $g\w{\ell}$. The following equations relate the conduction-electron Green's function to the desired spectral densities:
\begin{equation}\label{eq:112}
  \rho^S_{0}(\epsilon, T) =
  - \frac{1}{\pi N} \operatorname{Im}
  \sum_{kk'} \mathbb{G}^{S}_{kk'},
\end{equation}
and
\begin{equation}\label{eq:113}
  \rho^S_{1}(\epsilon, T) =
  -\dfrac{2\rho^2}{\pi N} \operatorname{Im}
  \sum_{kk'} \epsilon_{k} \epsilon_{k'} \mathbb{G}^{S}_{kk'}.
\end{equation}

To determine $\rho^S_{0}$, we must sum both sides of Eq.~(\ref{eq:110}) over the momenta $k$ and $k'$: \begin{align}
  \label{eq:114}
  \sum\w{k,k'} \mathbb{G}^S_{kk'}(\epsilon) = \sum\w{k}\mathbb{G}^0_k + \dfrac{V^2}{N} \Big(\sum\w{k}\mathbb{G}_k^0\Big)^2\mathbb{G}_d^S(\epsilon). 
\end{align}

Given Eq.~(\ref{eq:111}), it is straightforward to compute the sums
over momenta on the right-hand side of Eq.~(\ref{eq:114}). It results
that
\begin{align}
  \label{eq:115}
  \sum\w{k}\mathbb{G}_k^0 = -\pi\rho N i +\order{\rho\epsilon}.
\end{align}

At the low temperatures of interest, terms of $\order{\rho\epsilon}$
can be safely neglected. Substitution of the right-hand side of
Eq.~(\ref{eq:115}) for the sums on the right-hand side of
Eq.~(\ref{eq:114}) reduces the latter to the expression
\begin{align}
  \label{eq:116}
  \sum\w{k,k'} \mathbb{G}^S_{kk'}(\epsilon,T)
  = -\pi \rho N\Big(i + \pi \rho V^2 \mathbb{G}_d^S(\epsilon,T) \Big).
\end{align}

Comparison with Eq.~(\ref{eq:112}) then yields the following
expression for the $\phi\w{0}$ spectral function:
\begin{equation}\label{eq:117}
  \rho_{0}^S(\epsilon, T) = \rho - \pi\rho\Gamma \rho_{d}^{S}(\epsilon, T),
\end{equation}
where
\begin{align}
  \label{eq:118}
  \rho_d^S(\epsilon, T) = -\dfrac{1}{\pi}\operatorname{Im}\mathbb{G}_d^S(\epsilon,T)
\end{align}
is the dot-level spectral density.

Analogous algebra relates $\rho\w{1}$ to $\rho_d^S$. Multiplication of both sides of Eq.~(\ref{eq:110}) by $\epsilon\w{k}\epsilon\w{k'}$ followed by summation over both momenta yields the equality
\begin{align}\label{eq:119}
  \sum\w{k,k'}
  \epsilon\w{k}\epsilon\w{k'}\mathbb{G}^S_{kk'}(\epsilon)
  = \sum\w{k}\epsilon_k^2\mathbb{G}^0_{k}+ \dfrac{V^2}{N}
  \Big(\sum\w{k}\epsilon\w{k}\mathbb{G}_k^0\Big)^2\mathbb{G}_d^S(\epsilon). 
\end{align}

The first term on the right-hand side of Eq.~(\ref{eq:119}) is of
$\order{\rho\epsilon}$. The sum within parentheses in the second
factor is given by an equality analogous to Eq.~(\ref{eq:115}):
\begin{align}
  \label{eq:120}
  \sum\w{k}\epsilon\w{k}\mathbb{G}_k^0 = -N+\order{\rho \epsilon}.
\end{align}

Comparison with Eq.~(\ref{eq:113}), now yields the following expression for the $\phi\w{1}$ spectral density:
\begin{equation}\label{eq:121}
    \dfrac{\pi}{2}\rho_{1}^S(\epsilon, T) = \rho\Gamma \rho_{d}^{S}(\epsilon, T).
\end{equation}

The left-hand side of Eq.~(\ref{eq:121}) can now substituted for $\rho
\Gamma\rho_d^S(\epsilon, T)$ in the last term on the right-hand side of
Eq.~(\ref{eq:117}), with the result
\begin{align}\label{eq:122}
    \rho_{0}^S(\epsilon, T) = \rho - \frac{\pi^{2}}{2} \rho_{1}^S(\epsilon, T).
\end{align}

The universal moments $\mclu{j}{00}(T/\tk)$ ($j=0,2$) are the energy
moments $\mcl{j}(T/\tk)$ for the particle-hole symmetric model. In
other words, they can be obtained from by Eq.~\eqref{eq:63}
with $\rho_0^S$ substituted for $\rho\w0$, in the integrand on the
right-hand side:
\begin{align}
  \label{eq:123}
      \mclu{j}{00} \equiv \frac{2}{\rho h} \int \left(-\frac{\partial 
        f\w{\beta}}{\partial \epsilon} \right) 
    \rho_{0}^S(\epsilon, T)(\beta\epsilon)^j\,\dd\epsilon\qquad(j=0,2). 
\end{align}

Likewise, the moments $\mclu{j}{11}$ ($j=0,2$) are related to the $\phi\w1$ spectral density:
\begin{align}
  \label{eq:124}
        \mclu{j}{11} \equiv \frac{2}{\rho h} \int \left(-\frac{\partial 
        f\w{\beta}}{\partial \epsilon} \right) 
    \rho_{1}^S(\epsilon, T)(\beta\epsilon)^j\,\dd\epsilon\qquad(j=0,2). 
\end{align}

Substitution of the right-hand side of Eq.~(\ref{eq:122}) for
$\rho_0^S$ on the right-hand side of Eq.~(\ref{eq:123})~and
comparison with Eq.~(\ref{eq:124}) then leads to the expression
\begin{align}
  \label{eq:125}
  \mclu{j}{00}(T/\tk) =& \dfrac2{h}\rho\int \left(-\frac{\partial 
        f\w{\beta}}{\partial \epsilon}
                         \right)(\beta\epsilon)^j\,\dd{\epsilon}\ncr
                         &-\dfrac{\pi^2}2\mclu{j}{11}(T/\tk)\qquad(j=0,2).
\end{align}

The integral on the right-hand side of Eq.~(\ref{eq:124}) is
dimensionless. For $j=0$, it is unitary, and Eq.~\eqref{eq:83}
follows. For $j=2$, the integral equals $\pi^2/3$, and
Eq.~\eqref{eq:84} follows.
%section app 
%\bibliography{referencias} 

%

\end{document}